\documentclass[aip,jcp,a4paper]{revtex4-1}

\usepackage{graphicx}
\usepackage[sort&compress]{natbib}
\usepackage{amsfonts}
\usepackage{amsmath}
\usepackage{dcolumn}
\usepackage{color}
\usepackage{todonotes}

\begin{document}
\title{Using Fluorescence Detected Two-Dimensional Spectroscopy to Investigate Initial Exciton Delocalization Between Coupled Chromophores}

\author{Marco Schr\"oter}
\affiliation{Institute of Physics, University of Rostock, Albert Einstein Stra{\ss}e 23-24, 18059 Rostock, Germany}
\author{T\~{o}nu Pullerits}
\affiliation{Chemical Physics and NanoLund, Lund University, Box 124, 22100 Lund, Sweden.}
\author{Oliver K\"uhn}
\email{oliver.kuehn@uni-rostock.de}
\affiliation{Institute of Physics, University of Rostock, Albert Einstein Stra{\ss}e 23-24, 18059 Rostock, Germany}
\begin{abstract}
F\"orster theory describes electronic exciton energy migration in molecular assemblies as an incoherent hopping process between donor and acceptor molecules. The rate is expressed in terms of the overlap integral between donor fluorescence and acceptor absorption spectra. Typical time scales for systems like photosynthetic antennae are on the order of a few picoseconds. Prior to transfer it is assumed that the initially excited donor molecule has equilibrated with respect to the local environment. However, upon excitation and during the equilibration phase the state of the system needs to be described by the full density matrix, including  coherences between donor and acceptor states. While being intuitively clear, addressing this regime experimentally has been a challenge until the recently reported advances in Fluorescence Detected Two-Dimensional Spectroscopy (FD2DS).  Here, we demonstrate using fourth order perturbation theory, the conditions for the presence of donor-acceptor coherence induced  cross-peaks at zero waiting time between the first and the second pair of pulses. The approach is illustrated for a heterodimer model which facilitates an analytical solution.
\end{abstract}
\maketitle

\section{Introduction}
F\"orster theory is the corner-stone of the quantum mechanical modeling of resonant excitation energy transfer between chromophores in molecular assemblies.~\cite{may11} Originally developed to describe energy transfer between alike molecules in solution,~\cite{forster48_55} it has found widespread applications in nanoscale systems ranging from self-assembled aggregates to photosynthetic pigment-protein complexes.~\cite{scholes03_57,beljonne09_6583,sener07_15723} It reduces the transport problem to the  determination of the Golden Rule type transfer rate between a donor (D) and an acceptor (A), which can be expressed in terms of the overlap integral between the donor's emission and the acceptor's absorption spectrum. Originally, F\"orster theory had been developed to describe interacting molecules. To accommodate situations where the transfer occurs between pigment pools, where each pool is hosting a delocalized excitation possibly extending over several nanometers, standard F\"orster theory has been modified to include the interaction between collective dipoles.~\cite{sener11_518}

F\"orster theory is based on a number of assumptions, such as the weak coupling and Markovian limit.  This is also reflected in the typical  time scales, i.e. for local equilibration, $\tau_{\rm equi}$, and transfer, $\tau_{\rm trans}$. For $\tau_{\rm equi} \ll \tau_{\rm trans}$, the actual transfer starts from a thermalized D state. The populations of the D and A states follow from a  Pauli Master Equation, which defines an incoherent model for the energy flow in DA systems. However, going back to the derivation of the Pauli Master Equation, which starts from the Liouville-von Neumann equation for the reduced density operator,  one notices that coherences between D and A have been neglected, which is appropriate in the spirit of the time scale separation mentioned above. On the other hand, the initial (laser) excitation prepares an eigenstate or a superposition of delocalized eigenstates of the DA system. Whether or not this delocalization is  relevant depends, of course, on the relation between Coulomb and system-bath coupling. Being interested in cases where such a delocalization is relevant,  there must be a regime where coherence density matrix elements between D and A play a role. This raises the question how this transient effect can be observed and characterized or, in other words, what happens before the F\"orster regime sets in.

Ultrafast spectroscopy, particularly transient absorption~\cite{pullerits97_10560} and photon echo 2D spectroscopy (PE2DS)~\cite{jonas03_425,brixner05_625} is well suited for studies of sub-picosecond dynamics. However, precise measurements of the phenomena and processes at the timescales shorter than the pulse length used in experiment is a challenge. During the pulse overlap non-resonant signals from the environment (solvent or protein, e.g.) can significantly distort or even dominate the signal. Recent developments in incoherent action detected coherent multidimensional spectroscopies (for a comparison of various two-dimensional (2D) spectroscopies see Ref.~\citenum{fuller15_667}) has changed this situation. The key point is that the incoherent signal (e.g., fluorescence) from the sample and the environment can easily be separated.~\cite{mueller18_1964} Various incoherent action signals have been applied to measure coherent spectra. For instance, photoelectron emission microscopy was used in 2D nanoscopy, revealing localization of light by a rough metal surface.~\cite{aeschlimann15_663} Photocurrent detected 2D spectroscopy has provided valuable information about photoinduced processes in quantum well~\cite{nardin13_28617} and quantum dot based materials.~\cite{karki14_5869} Fluorescence detected 2D spectroscopy (FD2DS) was used to investigate the conformation of molecular dimer complexes~\cite{lott11_16521,tiwari18_} as well as photosynthetic antenna systems.~\cite{karki18_} In all these approaches four collinear laser pulses bring the system to an excited state, which can generate the incoherent signals as photocurrent or fluorescence. Such incoherent signals do not carry directionality of the phase matching. Instead phase cycling or phase modulation of the four pulses is used to separate the different signal contributions. In conventional PE2DS the signal is dispersed in a spectrometer, which directly provides the detection frequency of the 2D representation. In FD2DS the ''detection frequency'' is obtained by taking an additional Fourier transform over the time delay between the third and fourth pulse.

Due to the difference in experimental setup, PE2DS and FD2DS carry different information. The theory of PE2DS is well-established.~\cite{abramavicius09_2350} The different contributions to the signal (ground state bleach (GSB), stimulated emission (SE), and excited state absorption (ESA)) are usually analyzed in terms of double-sided Feynman diagrams.~\cite{mukamel95_} In particular, cross-peaks, where excitation and detection frequencies are different, are known to carry information about coherent couplings between transitions and, as a function of the population delay time, $T_2$, about population flow. However, assessing the coherent couplings is hampered by the above mentioned difficulties to access the $T_2\approx 0$ regime as well as the fact that GSB/SE and ESA contribute with different signs, which could lead to strong distortions or even cancellation of the signal. In passing we note that for larger population times the different contributions to PE2DS can be disentangled using polarized pulse sequences.~\cite{thyrhaug16_1653}

Recent application of FD2DS to the photosynthetic antenna complex LH2 of purple bacteria has revealed a cross-peak at $T_2=0$ indicating coherent coupling between the otherwise weakly interacting B800 and B850 pigment pools.~\cite{karki18_} Such a feature had not unequivocally been observed with PE2DS before (see, e.g., Refs.~\citenum{harel12_706} and \citenum{schroter18_1340}). In Ref. \citenum{karki18_} this was attributed to the fact that in LH2 rapid exciton-exciton annihilation leads to cancellation of ESA contributions to the FD2DS signal such that clean GSB can be observed at $T_2=0$. In the present contribution we aim to substantiate this argument by developing a perturbative expression for the FD2DS signal. In order to facilitate an analytical solution, the general formalism is specified to the case of a molecular heterodimer whose dynamics is described by means of a simple rate model. This model captures by no means the physics of LH2 or other light harvesting complexes with their complicated band structure but allows us to  demonstrate some general aspects of FD2DS applied to systems of coupled chromophores. The theoretical model is outlined in Section \ref{sec:theory}, starting with the Feynman diagram analysis of the fourth-order signal in Section \ref{sec:feynman}. Next the rate model is introduced and analytical expressions for the signal are given in Section \ref{eq:ratemodel} and discussed in Section \ref{sec:discussion}. Finally,  conclusions are presented in Section \ref{sec:conclusions}.
\section{Theoretical Model}%%%%%%%%%%%%%%%%%%%%%%%%%%%%%%%%%%%%%%%%%%%%%%%
\label{sec:theory}
\subsection{Feynman Diagrams for 4th Order Populations}
\label{sec:feynman}
\begin{figure}[th]
  	\centering
  	\includegraphics[width=0.5\textwidth]{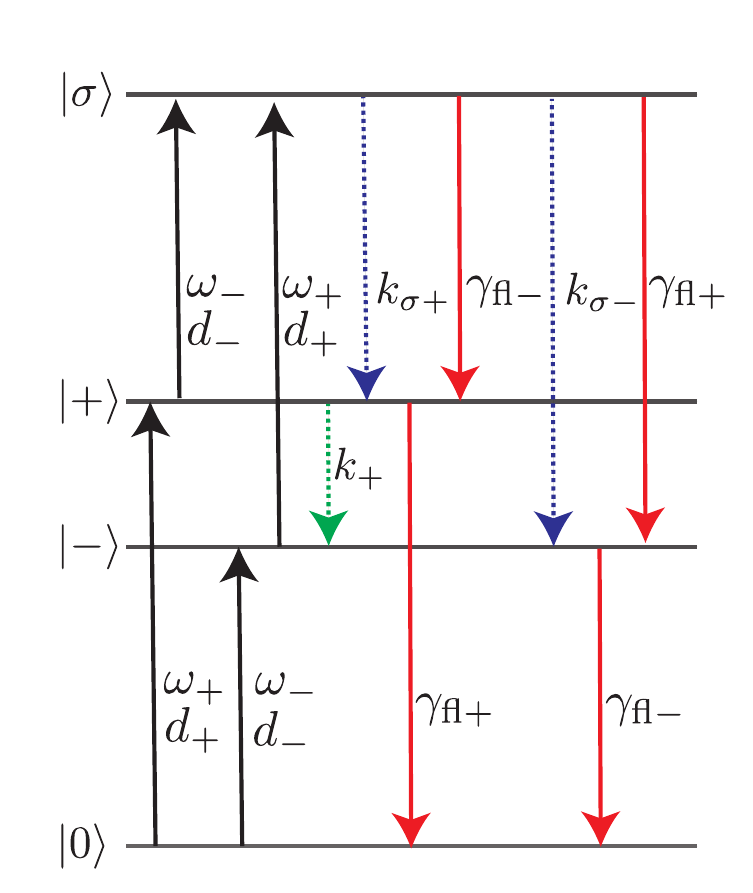}
  	\caption{Level scheme for heterodimer model with relaxation rates. For brevity we introduced $\omega_\pm = \omega_{\pm 0}=(E_\pm -E_0)/\hbar$ and $d_\pm = d_{\pm 0}$. Note that it holds $\omega_{\sigma \pm}=\omega_\mp$  as well as $d_{\sigma \pm}=d_\mp$ (assuming $d_A=d_D$).}
  	\label{fig:Level_Scheme}
 	\end{figure}

In the following we will consider the Frenkel exciton model of a heterodimer  with local  states $|D\rangle$ and $|A\rangle$ having excitation energies $E_D$ and $E_A$, respectively, and Coulomb coupling $J$. The difference in local excitation energies is given by $\Delta E=E_D-E_A$. The dimer is coupled via its transition dipole moments $d_{D0}\equiv d_D$ and $d_{A0} \equiv d_A$ to some external field $E(t)$.  The Hamiltonian in terms of the  one-exciton, $|\xi=\pm\rangle$ , and two-exciton, $|\sigma\rangle$, eigenstates is given by (cf. Fig.~\ref{fig:Level_Scheme})
	\begin{align}
		H&=\sum\limits_{\xi=\pm}E_\xi|\xi\rangle\langle \xi|+ E_\sigma|\sigma\rangle\langle \sigma|-E(t)d \, ,
  \end{align}
  with the dipole operator connecting ground and one-exciton states as well as one- and two-exciton states
  \begin{align}
		d&=\sum\limits_{\xi=\pm}\left(d_{\xi 0}|\xi\rangle\langle 0|+d_{\sigma\xi}|\sigma\rangle\langle \xi|\right)+ \mathrm{h.c.}
  \end{align}
  The one-exciton states can be expressed in terms of the local D and A states according to $ | \xi \rangle = C_{D}(\xi) |D \rangle +  C_{A}(\xi) |A \rangle$; the coefficients as well as the one-exciton energies are given in the Supplementary Material (Suppl. Mat.). For the heterodimer there is a single two-exciton state $ | \sigma \rangle = |D \rangle|A \rangle$ having energy $ E_\sigma=E_++E_-=E_D + E_A$. The transition dipole moments are given by
  \begin{align}
	 d_{\xi 0} &= C_D(\xi) d_D + C_A(\xi) d_A \, ,\\
	 d_{\sigma \xi}  &= C_D^*(\xi) d_A + C_A^*(\xi) d_D \, .
 \end{align}

  \begin{figure}[t]
   \includegraphics[width=0.6\columnwidth]{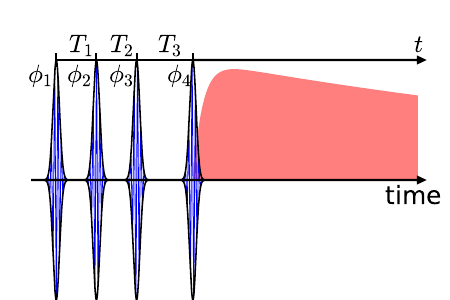}
   \caption{Pulse sequence and the  times characterizing the  maximum of the pulse envelope  corresponding to Eqs.~\eqref{eq:field_rephasing} and \eqref{eq:field_nonrephasing}. Also sketched is the fluorescence (red shaded area), which is integrated in the present detection scheme.}
   \label{fig:Pulse_Sequence}
  \end{figure}

In FD2DS experiments four laser pulses interact with the sample as  sketched in Fig.~\ref{fig:Pulse_Sequence}. The observable is the time-integrated fluorescence with any time information coming  solely from the timing of the four pulses. Thus it should not be confused with time-resolved fluorescence detection as studied, e.g., in Ref. \citenum{balevicius15_74101} for a weakly coupled heterodimer system. According to our model (cf. Fig.~\ref{fig:Level_Scheme}) the one-exciton states decay radiatively with rates $\gamma_{\rm fl\pm}$. For the coupled system and assuming that the energy splitting of the one-exciton states is much larger than the thermal energy, only  the  lower one-exciton state will be fluorescing. The equilibration between the one-exciton states, i.e. the transition $|+\rangle \rightarrow |-\rangle$,  shall proceed with a rate $k_+$. In addition the four pulse interactions can populate the two-exciton state, which can radiatively decay into the one-exciton states with rates $\gamma_{\rm fl \pm}$. Note that in principle there could be a direct two-photon emission leading to the ground state. This is, however, a rather unlikely process and therefore will be neglected. Competing with the radiative decay of the two-exciton state is the nonradiative deactivation via annihilation with rates $k_{\sigma \pm}$. The nonradiative decay of the two-exciton state is a consequence of the coupling to local doubly excited states, which rapidly decay via nonadiabatic transitions. The details of this coupling are strongly dependent on the energetic mismatch and the ratio of transition dipole moments for the delocalized two-exciton state and the local doubly excited state.~\cite{kuhn96_8586,kuhn97_809,renger97_3406,bruggemann03_746,yan12_105004} In passing we note that one can also view the process of annihilation as a sequence of exciton fusion and nonadiabatic deactivation at a certain chromophore.~\cite{may09_10086} In any case, to keep the model simple, we will discuss the two decay channels of the two-exciton state in terms of the ratio between the respective rates, $k_{\sigma \pm}$ and $\gamma_{\rm fl \pm}$, only.

In analogy to PE2DS the absorptive signal in rotating wave approximation can be split into a rephasing and a nonrephasing part, i.e.
\begin{align}
 S(\omega_1,T_2,\omega_3)	=& S^{\rm (R)}(-\omega_1,T_2,\omega_3)+S^{\rm (NR)}(\omega_1,T_2,\omega_3)
\end{align}
with
\begin{align}
  \label{eq:sig1}
 S^{\rm (R)}(-\omega_1,T_2,\omega_3)=& \int dT_1 dT_3 e^{-i\omega_1 T_1+i \omega_3 T_3} \int_0^\infty dt  {P}_{f}^{\rm (R)}(t,T_3,T_2,T_1)
\end{align}
and $S^{(\rm NR)}(\omega_1,T_2,\omega_3)$ alike. Here, we defined
\begin{align}
  {P}_{f}(t,T_3,T_2,T_1) &=  {P}_{f}^{\rm (R)}(t,T_3,T_2,T_1)+  {P}_{f}^{\rm (NR)}(t,T_3,T_2,T_1) \end{align}
with  ${P}_{f}(t)={P}_{f}(t,T_3,T_2,T_1)$ being the population of the fluorescent state $|f\rangle$ in fourth-order with respect to the incoming laser fields at detection time $t$. It depends parametrically on the delay times, $T_i$, of  the fields. Further, in Eq.~\eqref{eq:sig1} we assumed a time-integrated detection of the fluorescence.

\begin{figure*}[t]
		\centering
   \includegraphics[width=0.9\textwidth]{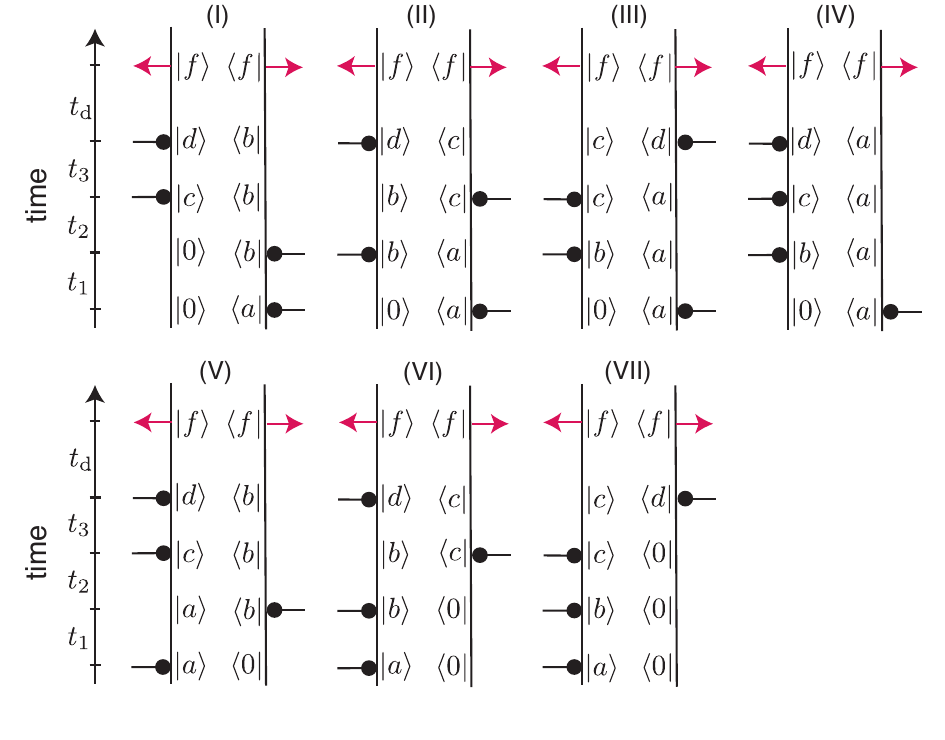}
   \caption{Feynman diagrams corresponding to Eq.~\eqref{eq:signal} (the full set of  diagrams is obtained by adding the respective hermitian conjugates). The horizontal lines with a dot can represent either and in- or outgoing arrow, depending on the actual interaction scheme (see Fig. \ref{fig:Feynmann_R}). The outgoing red arrows stand for the fluorescence from state $|f\rangle$. The labels $a,b,c,d$ denote any eigenstate $|\xi\rangle $ or $|\sigma \rangle$.}
   \label{fig:Feynmann_all}
  \end{figure*}

These experimental conditions have been previously studied by solving equations of motion including the external fields as well as their phase modulation explicitly.~\cite{damtie17_053830} Here, we will use an alternative approach, which is based on response functions.  Using fourth-order time-dependent perturbation theory one obtains (see Suppl. Mat.)

\begin{align}
\label{eq:signal}
	  {P}_{f}(t)	 = &\int_{0}^\infty  dt_{\mathrm{d}}  dt_3  dt_2  dt_1 \, E(t-t_{\mathrm{d}})E(t-t_{\mathrm{d}}-t_3)
	  E(t-t_{\mathrm{d}}-t_3-t_2)\nonumber\\ \times&  E(t-t_{\mathrm{d}}-t_3-t_2-t_1) {\mathcal R}_{f}(t_{\mathrm{d}},t_3,t_2,t_1) \, .
\end{align}
Here, we introduced the  fourth-order response function, ${\mathcal R}_{f}(t_{\mathrm{d}},t_3,t_2,t_1)$, which can be expressed in terms of 14 double~sided Feynman diagrams, see Fig.~\ref{fig:Feynmann_all}.

In order to proceed, we single out rephasing and non-rephasing contributions by defining the rephasing field
	\begin{align}
	 \label{eq:field_rephasing}
	 E(t)=\tilde{E}&\left\{\mathcal{E}(t+T_3+T_2+T_1)e^{i\omega_1 (t+T_3+T_2+T_1) + i\phi_1}+ \mathcal{E}(t+T_3+T_2)e^{-i\omega_2 (t+T_3+T_2) - i\phi_2}\right.\nonumber\\
	 &\left.+\mathcal{E}(t+T_3)e^{-i\omega_3 (t+T_3) - i\phi_3}+\mathcal{E}(t)e^{i\omega_4t + i\phi_4}\right\}
 	\end{align}
and the non-rephasing field
	\begin{align}
	 \label{eq:field_nonrephasing}
	 E(t)=\tilde{E}&\left\{\mathcal{E}(t+T_3+T_2+T_1)e^{-i\omega_1 (t+T_3+T_2+T_1) - i\phi_1}+ \mathcal{E}(t+T_3+T_2)e^{i\omega_2 (t+T_3+T_2) + i\phi_2}\right.\nonumber\\
	 &\left.+\mathcal{E}(t+T_3)e^{-i\omega_3 (t+T_3) - i\phi_3}+\mathcal{E}(t)e^{i\omega_4t + i\phi_4}\right\} \, .
 \end{align}
Here, the $\phi_i$ are the phases imprinted on the  fields and the $T_i$ are the times characterizing the  maximum of the pulse envelopes, ${\mathcal E}(t)$ (cf. Fig~\ref{fig:Pulse_Sequence}). The rephasing and nonrephasing contribution is detected at $\Phi_{\rm R}= \phi_1- \phi_2 - \phi_3 + \phi_4 $ and  $\Phi_{\rm NR}= -\phi_1+ \phi_2 - \phi_3 + \phi_4$, respectively.

\begin{figure*}[t]
  \includegraphics[width=0.9\textwidth]{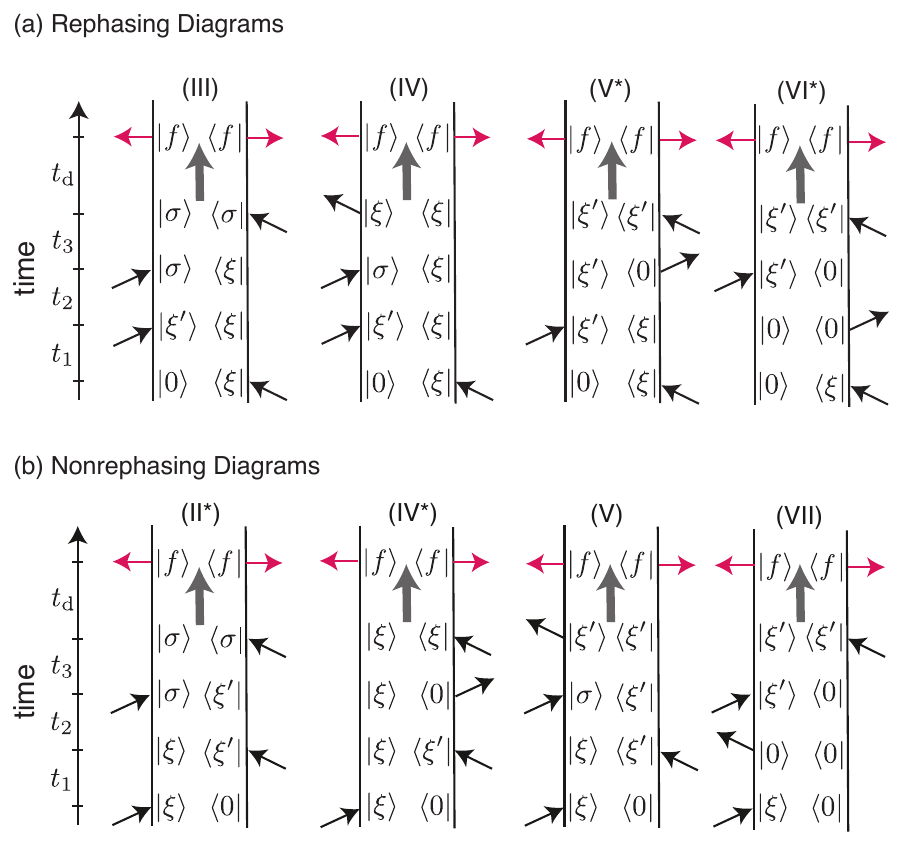}
  \caption{All fourth-order Feynman diagrams corresponding to the rephasing (a) and nonrephasing (b)  contribution to Eq.~\eqref{eq:R} and \eqref{eq:NR}, respectively,  for the heterodimer model. The grey arrow denotes a relaxation process from the state after the final field interaction to the fluorescent state. Note that diagrams (III), (IV), (II$^*$), and (V) can be classified as being of ESA type, whereas (V$^*$), (IV$^*$) are of SE type, and (VI$^*$) as well as (VII) are of GSB type.}
  \label{fig:Feynmann_R}
 \end{figure*}

Invoking the impulsive limit, ${\mathcal E}(t) \sim \delta(t)$, we obtain
 \begin{align}
	 {P}_{f}^{\rm (R)}(t_{\mathrm{d}},t_3,t_2,t_1)&=\tilde{E}^4 e^{i\Phi_{\rm R}}\,{\mathcal R}_{f}^{\rm (R)}(t_{\mathrm{d}},T_3,T_2,T_1)\, ,\\
	  {P}_{f}^{\rm (NR)}(t_{\mathrm{d}},t_3,t_2,t_1)&=\tilde{E}^4 e^{i\Phi_{\rm NR}}\,{\mathcal R}_{f}^{\rm (NR)}(t_{\mathrm{d}},T_3,T_2,T_1)\, .
	 \end{align}
The response functions for the heterodimer model are given as (note that in the impulsive limit one has $T_i=t_i$ and $t=t_{\rm d}$)

\begin{align}
\label{eq:R}
\mathcal{R}_{f}^{\rm (R)}(t_{\mathrm{d}},t_3,t_2,t_1)	=&	R_{f}^{({\rm III})}(t_{\mathrm{d}},t_3,t_2,t_1)+R_{f}^{({\rm IV})}(t_{\mathrm{d}},t_3,t_2,t_1)\nonumber\\
&+R_{f}^{({\rm V^*})}(t_{\mathrm{d}},t_3,t_2,t_1)+R_{f}^{({\rm VI^*})}(t_{\mathrm{d}},t_3,t_2,t_1)\, ,
\end{align}
\begin{align}
  \label{eq:NR}
	 \mathcal{R}_{f}^{\rm (NR)}(t_{\mathrm{d}},t_3,t_2,t_1)	=&	R_{f}^{({\rm II^*})}(t_{\mathrm{d}},t_3,t_2,t_1)+R_{j}^{({\rm IV^*})}(t_{\mathrm{d}},t_3,t_2,t_1)\nonumber\\
&	+R_{f}^{({\rm V})}(t_{\mathrm{d}},t_3,t_2,t_1)+R_{f}^{({\rm VII})}(t_{\mathrm{d}},t_3,t_2,t_1)
 \end{align}
and presented in terms of Feynman diagrams in Fig.~\ref{fig:Feynmann_R}.

Note that the FD2DS detection is not sensitive to coherences generated by the last pulse, i.e. after the four interactions all diagrams end in a population density matrix element.

\subsection{Rate Model Analysis}
\label{eq:ratemodel}
Next, the general model developed in the previous subsection is specified to the  simple rate model describing the radiative and non-radiative population flow in the heterodimer according to Fig.~\ref{fig:Level_Scheme}. In addition, we  will assume a simple lineshape model, where the one-exciton and two-exciton transitions are homogeneously broadened by a rate $\gamma_\pm$ and $\gamma_{\sigma \pm}$, respectively. Thus, for the rephasing contributions one obtains
\begin{align}
	 R_{f}^{({\rm III})}(t_{\rm d},t_3,t_2,t_1) =&
	 \sum_{\xi,\xi',\sigma} d_\xi d_{\xi'} d_{\sigma\xi'} d_{\xi \sigma}\,
\mathcal{P}_{\sigma\rightarrow f}(t_{\rm d})
e^{-i\omega_{\sigma \xi} t_3 - \gamma_{\sigma\xi} t_3}
e^{i\omega_{\xi} t_1 -\gamma_\xi t_1}									\nonumber\\
& \qquad\times e^{-\delta_{\xi,\xi'}k_\xi t_2- (1-\delta_{\xi,\xi'})(i\omega_{\xi \xi'}+\gamma_{\xi\xi'}) t_2} \, ,\\
	 R_{f}^{({\rm IV})}(t_{\rm d},t_3,t_2,t_1) 	=&	-\sum_{\xi,\xi',\sigma} d_\xi d_{\xi'} d_{\sigma\xi'}d_{\sigma\xi}\,
\mathcal{P}_{\xi\rightarrow f}(t_{\rm d})
e^{-i\omega_{\sigma \xi} t_3 -\gamma_{\sigma\xi}t_3}
e^{i\omega_{\xi} t_1 -\gamma_\xi t_1}									\nonumber\\
&	\quad\qquad\times  e^{-\delta_{\xi,\xi'}k_\xi t_2 - (1-\delta_{\xi,\xi'})(i\omega_{\xi \xi'}+\gamma_{\xi\xi'}) t_2 }  \, , \\
	 R_{f}^{({\rm V^*})}(t_{\rm d},t_3,t_2,t_1) =&	-\sum_{\xi, \xi'} |d_{\xi'}|^2 |d_\xi|^2\,
\mathcal{P}_{\xi'\rightarrow f}(t_{\rm d})
e^{ - i\omega_{\xi'} t_3-\gamma_{\xi'}t_3}
e^{i\omega_{\xi } t_1 - \gamma_{\xi}t_1}							\nonumber\\
&	\quad\qquad\times e^{-\delta_{\xi,\xi'}k_\xi t_2 - (1-\delta_{\xi,\xi'})(i\omega_{\xi \xi'}+\gamma_{\xi\xi'}) t_2 }  \, , \\
	 R_{f}^{({\rm VI^*})}(t_{\rm d},t_3,t_2,t_1)=&	-\sum_{\xi, \xi'} |d_{\xi'}|^2 |d_\xi|^2\,
\mathcal{P}_{\xi'\rightarrow f}(t_{\rm d})
e^{ - i\omega_{\xi'} t_3-\gamma_{\xi'}t_3}
e^{i\omega_{\xi } t_1 - \gamma_{\xi}t_1} \, .
 \end{align}
 The non-rephasing contributions read
\begin{align}
	 R_f^{({\rm II}^*)}(t_{\rm d},t_3,t_2,t_1)=& 	\sum_{\xi,\xi',\sigma} d_\xi d_{\xi'} d_{\sigma\xi'} d_{\xi \sigma}\,
\mathcal{P}_{\sigma\rightarrow f}(t_{\rm d})
e^{-i\omega_{\sigma \xi'} t_3 - \gamma_{\sigma\xi'} t_3}
e^{-i\omega_{\xi} t_1 -\gamma_\xi t_1}									\nonumber\\
& 	\times e^{-\delta_{\xi,\xi'}k_\xi t_2- (1-\delta_{\xi,\xi'})(-i\omega_{\xi \xi'}+\gamma_{\xi'\xi}) t_2}  \, , \\
	 R_f^{({\rm IV^*})}(t_{\rm d},t_3,t_2,t_1)=& -	\sum_{\xi,\xi'} |d_\xi|^2 |d_{\xi'}|^2\,
\mathcal{P}_{\xi\rightarrow f}(t_{\rm d})
e^{-i\omega_{\xi} t_3 -\gamma_\xi t_3}
e^{-i\omega_{\xi} t_1 -\gamma_\xi t_1}									\nonumber\\
&	\times  e^{-\delta_{\xi,\xi'}k_\xi t_2 - (1-\delta_{\xi,\xi'})(-i\omega_{\xi \xi'}+\gamma_{\xi'\xi}) t_2 }   \, ,\\
	 R_f^{({\rm V})}(t_{\rm d},t_3,t_2,t_1)		=& -	\sum_{\xi,\xi',\sigma} d_\xi d_{\xi'} d_{\sigma\xi'}d_{\sigma\xi}\,
\mathcal{P}_{\xi'\rightarrow f}(t_{\rm d})
e^{-i\omega_{\sigma \xi'} t_3 -\gamma_{\sigma\xi}t_3}
e^{-i\omega_{\xi} t_1 -\gamma_\xi t_1}									\nonumber\\
&	\times  e^{-\delta_{\xi,\xi'}k_\xi t_2 - (1-\delta_{\xi,\xi'})(-i\omega_{\xi \xi'}+\gamma_{\xi'\xi}) t_2 }  \, , \\
	 R_f^{({\rm VII})}(t_{\rm d},t_3,t_2,t_1) =& - \sum_{\xi, \xi'} |d_{\xi'}|^2 |d_\xi|^2\,
\mathcal{P}_{\xi'\rightarrow f}(t_{\rm d})
e^{-i\omega_{\xi'} t_3-\gamma_{\xi'}t_3}
e^{-i\omega_{\xi} t_1-\gamma_{\xi}t_1}\, .
 \end{align}
 Here, we introduced the functions $\mathcal{P}_{\sigma\rightarrow f}(t_{\rm d})$ and
 $\mathcal{P}_{\xi \rightarrow f}(t_{\rm d})$, which describe the relaxation of the two-exciton and one-exciton state, respectively, to the fluorescent state $f$ (grey arrows in Fig. \ref{fig:Feynmann_R}). Specific expressions can be obtained by considering the  rate model according to Fig.~\ref{fig:Level_Scheme}
	\begin{align}
    \label{eq:rate-eq1}
	 \frac{dP_\sigma}{dt} =& - (\Gamma_{\sigma +}+\Gamma_{\sigma -}) P_\sigma \, , \\
	 \frac{dP_+}{dt} =& \Gamma_{\sigma +} P_\sigma - \Gamma_+ P_+ \, ,\\
	 \frac{dP_-}{dt} =& \Gamma_{\sigma -} P_\sigma + k_+ P_+ -\gamma_{\mathrm{fl-}} P_- \, ,
   \label{eq:rate-eq3}
 \end{align}
with $P_\pm$ and $P_\sigma$ being the populations of the respective states.
Further, $\Gamma_{\sigma \pm} = k_{\sigma \pm}+\gamma_{\mathrm{fl}\mp}$ denote the sum of nonradiative ($k_{\sigma \pm}$) and radiative ($\gamma_{\mathrm{fl}\pm}$) decay rates of state $|\sigma\rangle $ to state $|\pm\rangle$ and $\Gamma_+=k_+ + \gamma_{\rm fl+}$ the sum of the decay rates of state $|+\rangle$.

 The system of rate equations (\ref{eq:rate-eq1}-\ref{eq:rate-eq3}) can be solved analytically. Thereby, the initial conditions have to be chosen according to the considered processes, i.e. $\mathcal{P}_{\sigma\rightarrow f}(t_{\rm d})$ and
 $\mathcal{P}_{\xi \rightarrow f}(t_{\rm d})$. Specifically, we obtain for the three fluorescence channels $f=(-,+,\sigma)$
 \begin{align}
  \label{eq:rate-signal-contributions}
 	\mathcal{P}_{-\rightarrow -}(t_{\mathrm{d}})=P_-(t) \quad\text{for}\quad P_\sigma(0)=0, P_+(0)=0, P_-(0)=1 \, ,\nonumber \\
  \mathcal{P}_{+\rightarrow +}(t_{\mathrm{d}})=P_+(t) \quad\text{for}\quad P_\sigma(0)=0, P_+(0)=1, P_-(0)=0\, ,\nonumber \\
	\mathcal{P}_{+\rightarrow -}(t_{\mathrm{d}})=P_-(t) \quad\text{for}\quad P_\sigma(0)=0, P_+(0)=1, P_-(0)=0\, ,\nonumber \\
  \mathcal{P}_{\sigma\rightarrow \sigma}(t_{\mathrm{d}})=P_\sigma(t) \quad\text{for}\quad P_\sigma(0)=1, P_+(0)=0, P_-(0)=0\, ,\nonumber \\
  \mathcal{P}_{\sigma\rightarrow +}(t_{\mathrm{d}})=P_+(t) \quad\text{for}\quad P_\sigma(0)=1, P_+(0)=0, P_-(0)=0\, ,\nonumber \\
	\mathcal{P}_{\sigma\rightarrow -}(t_{\mathrm{d}})=P_-(t) \quad\text{for}\quad P_\sigma(0)=1, P_+(0)=0, P_-(0)=0\nonumber\, .
\end{align}
The analytical solutions of Eqs. (\ref{eq:rate-eq1}-\ref{eq:rate-eq3}) subject to these initial conditions are given in the Suppl. Mat.

The 2D spectrum features two diagonal peaks, $(\omega_1,\omega_3)=(\pm,\pm)$,  as well as two off-diagonal cross-peaks  $(\omega_1,\omega_3)=(\pm,\mp)$. The contributions to these peaks can be calculated using  the general Feynman diagrams in Fig.~\ref{fig:Feynmann_R}.  In the following, we only consider the case $t_2=0$ (zero waiting time) for simplicity.

The derivation of the signal will be sketched for the cross-peak at $(\omega_1,\omega_3)=(+,-)$; expressions for the other peaks are given in the Suppl. Mat. Using   $\omega_{\sigma+}=\omega_-$ we obtain for the case $f=-$
\begin{align}
 \mathcal{R}_-^{\rm (R)}(t_\mathrm{d},t_3,t_2=0,t_1)     =& e^{-i(-\omega_{+} t_1 +\omega_- t_3)}
 \left\{    -e^{-\gamma_- t_3 -\gamma_+ t_1 }|d_+|^2 |d_-|^2
\mathcal{P}_{-\rightarrow -}(t_\mathrm{d}) \right.\nonumber\\
 &                        -e^{-\gamma_- t_3 -\gamma_+ t_1 }|d_+|^2 |d_-|^2
\mathcal{P}_{-\rightarrow -}(t_\mathrm{d})\nonumber\\
 &                        +e^{-\gamma_{\sigma+} t_3 -\gamma_+ t_1 }|d_+|^2 |d_{\sigma+}|^2
 \mathcal{P}_{\sigma\rightarrow -}(t_\mathrm{d})\nonumber\\
 &                        -e^{-\gamma_{\sigma+} t_3 -\gamma_+ t_1 }|d_+|^2 |d_{\sigma+}|^2
 \mathcal{P}_{+\rightarrow -}(t_\mathrm{d})\nonumber\\
 &                        +e^{-\gamma_{\sigma+} t_3 -\gamma_+ t_1 }d_+ d_- d_{\sigma+} d_{\sigma-}
 \mathcal{P}_{\sigma\rightarrow -}(t_\mathrm{d})\nonumber\\
 &        \left.    -e^{-\gamma_{\sigma+} t_3 -\gamma_+ t_1 }d_+ d_- d_{\sigma+} d_{\sigma-}
 \mathcal{P}_{+\rightarrow -}(t_\mathrm{d})\right\}\, ,
 \end{align}
 \begin{align}
 \mathcal{R}_-^{\rm (NR)}(t_\mathrm{d},t_3,t_2=0,t_1)     =& e^{-i(\omega_{+} t_1 +\omega_- t_3)}
 \left\{    -e^{-\gamma_- t_3 -\gamma_+ t_1 } |d_{+}|^2 |d_-|^2
 \mathcal{P}_{-\rightarrow -}(t_\mathrm{d})\right.\nonumber\\
 &                        +e^{-\gamma_{\sigma+} t_3 -\gamma_+ t_1 }|d_+|^2 |d_{\sigma+}|^2
 \mathcal{P}_{\sigma\rightarrow -}(t_\mathrm{d})\nonumber\\
 &                        -e^{-\gamma_{\sigma+} t_3 -\gamma_+ t_1 }|d_+|^2 |d_{\sigma+}|^2
 \left.        \mathcal{P}_{+\rightarrow -}(t_\mathrm{d})\right\}\, ,
\end{align}
and for the case $f=+$
\begin{align}
 \mathcal{R}_+^{\rm (R)}(t_\mathrm{d},t_3,t_2=0,t_1)     =& e^{-i(-\omega_{+} t_1 +\omega_- t_3)}
 \left\{e^{-\gamma_{\sigma+} t_3 -\gamma_+ t_1 }|d_+|^2 |d_{\sigma+}|^2
   \mathcal{P}_{\sigma\rightarrow +}(t_\mathrm{d})\right.\nonumber\\
   &                        -e^{-\gamma_{\sigma+} t_3 -\gamma_+ t_1 }|d_+|^2 |d_{\sigma+}|^2
   \mathcal{P}_{+\rightarrow +}(t_\mathrm{d})\nonumber\\
   &           -e^{-\gamma_{\sigma+} t_3 -\gamma_+ t_1 }d_+ d_- d_{\sigma+} d_{\sigma-}
   \mathcal{P}_{+\rightarrow +}(t_\mathrm{d})\nonumber \\
 &\left.+e^{-\gamma_{\sigma+} t_3 -\gamma_+ t_1 }d_+ d_- d_{\sigma+} d_{\sigma-}
  \mathcal{P}_{\sigma\rightarrow +}(t_\mathrm{d})\right\}\, ,
\end{align}
\begin{align}
  \mathcal{R}_+^{\rm (NR)}(t_\mathrm{d},t_3,t_2=0,t_1)     =& e^{-i(\omega_{+} t_1 +\omega_- t_3)}
  \left\{e^{-\gamma_{\sigma+} t_3 -\gamma_+ t_1 }|d_+|^2 |d_{\sigma+}|^2
    \mathcal{P}_{\sigma\rightarrow +}(t_\mathrm{d})\right. \nonumber \\
    &                        -e^{-\gamma_{\sigma+} t_3 -\gamma_+ t_1 }|d_+|^2 |d_{\sigma+}|^2
    \left.        \mathcal{P}_{+\rightarrow +}(t_\mathrm{d})\right\}\, ,
\end{align}

and, finally, for $f=\sigma$
\begin{align}
 \mathcal{R}_\sigma^{\rm (R)}(t_\mathrm{d},t_3,t_2=0,t_1)     =& e^{-i(-\omega_{+} t_1 +\omega_- t_3)}
 \left\{e^{-\gamma_{\sigma+} t_3 -\gamma_+ t_1 }|d_+|^2 |d_{\sigma+}|^2
   \mathcal{P}_{\sigma\rightarrow \sigma}(t_\mathrm{d})\right.\nonumber\\
 &\left.+e^{-\gamma_{\sigma+} t_3 -\gamma_+ t_1 }d_+ d_- d_{\sigma+} d_{\sigma-}
  \mathcal{P}_{\sigma\rightarrow \sigma}(t_\mathrm{d})\right\}\, ,
\end{align}
\begin{align}
  \mathcal{R}_\sigma^{\rm (NR)}(t_\mathrm{d},t_3,t_2=0,t_1)     =& e^{-i(\omega_{+} t_1 +\omega_- t_3)}
  e^{-\gamma_{\sigma+} t_3 -\gamma_+ t_1 }|d_+|^2 |d_{\sigma+}|^2
    \mathcal{P}_{\sigma\rightarrow \sigma}(t_\mathrm{d})\, .
\end{align}
The signal is obtained after insertion of the solutions of the rate equations, integration with respect to the detection type, and taking the Fourier transform. Introducing Lorentzian type lineshape functions for the real part of the absorptive 2D spectra and assuming for simplicity $\gamma_{\sigma+}=\gamma_+$ and $\gamma_{\sigma-}=\gamma_-$ yields

\begin{align}
  \label{eq:ss1}
  &\tilde{S}(\omega_1,\omega_3) = {\rm Re}[ S(\omega_1,T_2=0,\omega_3)]=\nonumber\\
  & - |d_+|^2 |d_-|^2 \left( 2 L_{+,-}(\omega_1,\omega_3) + L^{\rm (R)}_{+,-}(\omega_1,\omega_3)  \right) \int\limits_0^\infty dt  \mathcal{P}_{-\rightarrow-}(t) \nonumber\\
  & + \left( |d_+|^2 |d_{\sigma+}|^2 L_{+,-}(\omega_1,\omega_3)+ d_+ d_{\sigma+} d_- d_{\sigma-} L^{\rm (R)}_{+,-}(\omega_1,\omega_3) \right) \nonumber\\
  &\quad\times\int\limits_0^\infty dt  (2\mathcal{P}_{\sigma\rightarrow\sigma}(t)+\mathcal{P}_{\sigma\rightarrow+}(t)+\mathcal{P}_{\sigma\rightarrow-}(t)-\mathcal{P}_{+\rightarrow+}(t)-\mathcal{P}_{+\rightarrow-}(t)) \, ,
\end{align}
with
\begin{align}
 	L^{\rm (NR)}_{a,b}(\omega_1,\omega_3)=&\mathrm{Re}\int dT_1 dT_3 e^{i(\omega_1-\omega_a) T_1+(i\omega_3-\omega_b) T_3} e^{ -\gamma_a T_1 -\gamma_b T_3}\nonumber\\
 	=&\frac{-(\omega_1-\omega_a)(\omega_3-\omega_b)+\gamma_a\gamma_b} {[(\omega_1-\omega_a)^2+\gamma_a^2][(\omega_3-\omega_b)^2+\gamma_b^2]}\, ,\\
 	L^{\rm (R)}_{a,b}(\omega_1,\omega_3)=&\mathrm{Re}\int dT_1 dT_3 e^{i(-\omega_1+\omega_a) T_1+i(\omega_3-\omega_b) T_3} e^{-\gamma_a T_1 -\gamma_b T_3 }\nonumber\\
	=&\frac{(\omega_1-\omega_a)(\omega_3-\omega_b)+\gamma_a\gamma_b}{[(\omega_1-\omega_a)^2+\gamma_a^2][(\omega_3-\omega_b)^2+\gamma_b^2]}\, ,\\
 	L_{a,b}(\omega_1,\omega_3)=&L^{\rm (R)}_{a,b}(\omega_1,\omega_3)+L^{\rm (NR)}_{a,b}(\omega_1,\omega_3)\nonumber\\
	=&\frac{2\gamma_a\gamma_b}{[(\omega_1-\omega_a)^2+\gamma_a^2][(\omega_3-\omega_b)^2+\gamma_b^2]}\, .
\end{align}

Note the appearance of the term $2\mathcal{P}_{\sigma\rightarrow\sigma}(t)$ in Eq.~\eqref{eq:ss1}, where the factor two accounts for the emission of two photons. Within our model this factor is due to the separation of the  pathways for diagrams III and II$^*$ according to the different final states.

In general the fluorescence lifetime for typical chromophores is on the order of a few nanoseconds, whereas the intraband relaxation between one-exciton states is of the order of some hundreds of femtoseconds to a few picoseconds.~\cite{pullerits96_381,kuhn97_4154}  The excited state absorption often features a broad band in the range of the two-exciton states.~\cite{ambrosek11_17649,kosumi11_92} Given an appreciable density of states  there should be  always a coupling between the two types of transitions. As a consequence  the nonradiative, i.e. annihilation, rate is large, leading to time scales of  a few hundred femtoseconds.~\cite{bruggemann09_140,hader17_31989,dostal18_2466} Hence, for coupled chromophores we can assume that $ k_{\sigma \pm} \simeq k_+ \gg \gamma_{\mathrm{fl \pm}}$ holds. In any case the experimental detection time is of the order of tens of nanoseconds, which justifies the upper integration limit   $t\rightarrow\infty$ in Eq.~\eqref{eq:sig1}.

 	The total signal in the limit of fast annihilation and rapid transfer, $k_{\sigma \pm} \simeq k_+ \gg \gamma_{\mathrm{fl} \pm}$ yields $\int\limits_0^\infty dt \mathcal{P}_{\sigma\rightarrow\sigma}(t) , \int\limits_0^\infty dt \mathcal{P}_{\sigma\rightarrow+}(t) , \int\limits_0^\infty dt\mathcal{P}_{+\rightarrow+}(t) \ll \int\limits_0^\infty dt \mathcal{P}_{+\rightarrow-}(t) = \int\limits_0^\infty dt\mathcal{P}_{\sigma\rightarrow-}(t)$ (see Suppl. Mat.). Thus, the second term in Eq. \eqref{eq:ss1}, which is due to ESA is negligible compared to the first term.  The signal is solely due to GSB/SE contributions (first term) and reads

	\begin{align}
	\label{eq:final1}
		\tilde{S}(\omega_1,\omega_3)=&-\frac{1}{\gamma_\mathrm{fl-}}\Big\{ 2 |d_-|^4 L_{-,-}(\omega_1,\omega_3)+ |d_-|^2 |d_+|^2 L^{\rm (NR)}_{-,-}(\omega_1,\omega_3)  \nonumber\\
		& +2|d_+|^4 L_{+,+}(\omega_1,\omega_3)+|d_-|^2 |d_+|^2 L^{\rm (NR)}_{+,+}(\omega_1,\omega_3)  \nonumber\\
		& +|d_-|^2 |d_+|^2 (L_{+,-}(\omega_1,\omega_3)+ L^{\rm (R)}_{+,-}(-\omega_1,\omega_3) ) \nonumber\\
		& +|d_-|^2 |d_+|^2 (L_{-,+}(\omega_1,\omega_3)+L^{\rm (R)}_{-,+}(-\omega_1,\omega_3) )\Big\}\, .
	\end{align}
However, in case of no coupling ($J=0$), the one-exciton states decouple ($k_+=0$) and there is no mixing between two-exciton state and the local doubly excited states such that $k_{\sigma \pm}=0$. In this case the only deactivation channel for the two-exciton state is fluorescence with rates $\gamma_{\rm fl \pm}$. Hence, all time integrals (ESA, GSB and SE) contribute to Eq. \eqref{eq:ss1}. In this case contributions to cross-peaks with different sign cancels out exactly (see Suppl. Mat.).
The signal has only diagonal contributions given by

\begin{align}
  \label{eq:final2}
  \tilde{S}(\omega_1,\omega_3)=&
    -\frac{2|d_-|^4 L_{-,-}(\omega_1,\omega_3)}{\gamma_\mathrm{fl-}}-\frac{2|d_+|^4 L_{+,+}(\omega_1,\omega_3)}{\gamma_\mathrm{fl+}}\, .
\end{align}
Equations \eqref{eq:final1} and \eqref{eq:final2} constitute the main result of this paper.

\section{Discussion}	%%%%%%%%%%%%%%%%%%%%%%%%%%%%%%%%%%%%%%%%%%%%%%%%
\begin{figure}[t]
	  \centering
	  \includegraphics[width=0.5\columnwidth]{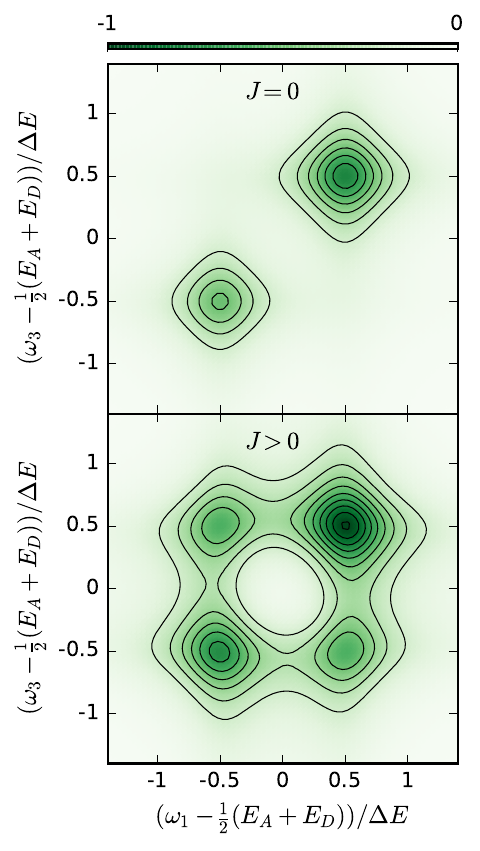}
	  \caption{FD2DS signal (in arb. units) according to Eqs.~\eqref{eq:final1} and \eqref{eq:final2} for $d_A=d_D$  and $\gamma_+=\gamma_-=0.3 \Delta E$. Upper panel $J=0$, lower panel $2J/\Delta E=0.1$.}
	  \label{fig:spectra}
\end{figure}

\label{sec:discussion}
An overview on the FD2DS signal  is provided in Fig.~\ref{fig:spectra} for cases with and without Coulomb coupling. We have chosen a H-dimer configuration ($J>0$), for a J-dimer ($J<0$) the spectrum has to be mirrored at the anti-diagonal (cf. Fig.~\ref{fig:dip} below). The  parameter $2J/\Delta E=0.1$ has been chosen only for the sake of having clearly separated peaks. However, the situation can be considered as being representative for various systems ranging from  the B800 and B850 pigment pools in LH2~\cite{karki18_} to chromophore dyads such as studied in Ref.~\citenum{tiwari18_}. Of course, the simple heterodimer is  far from mimicking the real LH2 with its intricate multilevel structure.~\cite{kuhn97_3432} Remarkably, the spectrum at $J\ne 0$ resembles   the one reported in Ref.~\citenum{karki18_}. Most notably, there are clear cross-peaks, which can be attributed to the effect of  coupling of the local transitions.

Using Eq.~\eqref{eq:final1} and assuming equal broadenings the dependence of the peak heights on the coupling strength can be addressed in more detail. The diagonal peaks scale like $2|d_\pm|^4+|d_+|^2|d_-|^2$ whereas the cross-peaks depend on $|d_+|^2|d_-|^2$. Assuming equal monomeric transition strengths, the dependencies of these peak amplitudes on the coupling strength are shown in Fig. \ref{fig:dip}. In case of a H-dimer the signal at the  lower diagonal diminishes whereas that of the upper diagonal increases with coupling strength. For a J-type dimer the situation is just the opposite. Inspecting the cross-peaks we notice that their intensity is equal and decreases with coupling strength, independent on the sign of $J$. This is a consequence of the fact that the Feynman diagrams involve a pathway via a state, which becomes increasingly dark with  stronger coupling. As a note in caution we should emphasize that the limit of a perfect H-dimer $2J/\Delta E\rightarrow \infty$ is not covered by the present model. In this case one has $|d_-|^2 \rightarrow 0$ and thus no fluorescence. However, the focus of the present work is on the limit of weak coupling and thus the perfect H-dimer case is not considered further on.

\begin{figure}[t]
	  \centering
	  \includegraphics[width=0.5\columnwidth]{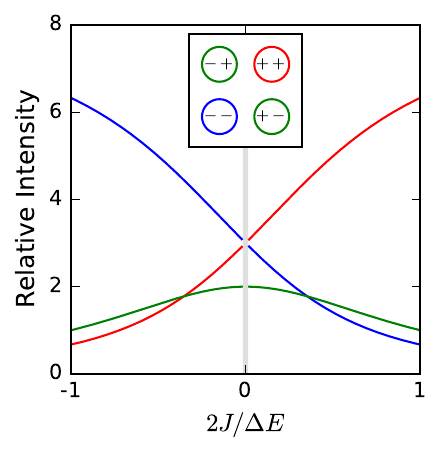}
	  \caption{Dependence of diagonal and off-diagonal peaks in the signal according to Eq. \eqref{eq:final1}  for $d_A=d_D$ on the  Coulomb coupling. The grey area indicates the fact that the model leading to Eq.~\eqref{eq:final1} breaks down for $2J/\Delta E \rightarrow 0$.}
	  \label{fig:dip}
\end{figure}
As noted above the case of $J=0$ requires special attention. According to Eq. \eqref{eq:final2} the diagonal peaks scale like $|d_\pm|^4$, i.e. similar to the case of non-vanishing coupling. As can be seen in Fig. \ref{fig:spectra} the cross-peaks carry no intensity due to cancellation of different Feynman diagrams. In Fig. \ref{fig:dip} the limit $J\rightarrow 0$ is not included since our model does not provide a continuous description. The latter would require to take into account the mixing between two-exciton states and localized double excitations as well as a description of the rate for nonadiabatic transitions,~\cite{may09_10086} which is beyond the scope of the present approach.

\section{Conclusions}
\label{sec:conclusions}
In the present contribution, we have developed a response function approach, which in combination with a simple rate model for population flow, yielded analytical expression for FD2DS signals. It has been shown that FD2DS is capable  of revealing the initially delocalized excitation of  coupled chromophores. Thus, FD2DS gives a means to address the dynamical regime \textit{before} incoherent F\"orster transfer sets in. The key point is that cross-peaks at population time zero are determined by GSB contributions only, which is in contrast to PE2DS. In order to demonstrate this, the developed formalism has been applied to a heterodimer system, which allowed to study the dependence of the signal on the Coulomb coupling in detail. Interestingly, since for the heterodimer the amplitude of the cross-peaks is proportional the product of the absolute values of the transition dipole moments of the upper and lower exciton state, it decreases with increasing coupling strength. This result must be considered as being special and a consequence of the high symmetry of the model system.

There is a wealth of systems in natural and artificial photosynthesis, where exciton transport between pigment pools is believed to occur by means of the F\"orster mechanism.~\cite{sener07_15723,sener11_518,beljonne09_6583, woller13_2759,kramer18_} There is no point in challenging this well-established picture. However, the initially delocalized excitation could have a functional role such as to direct excitation energy flow between weakly coupled pigment pools.~\cite{caycedo-soler17_6015} In order to better understand and possibly exploit initial delocalization in artificial light-harvesting, FD2DS could become an indispensable tool.

While the developed response function approach is rather general, there is plenty of room for improvement as far as the actual dynamics is concerned. On the level of a simple rate model, one could include the $J\rightarrow 0$ limit. This would require to incorporate local doubly excited state and respective rates for exciton fusion and internal conversion.~\cite{may09_10086,bruggemann01_11391} On a more elaborate level, the response function could be obtained by direct propagation using, e.g., multilevel Redfield theory~\cite{kuhn03_2155} or the more sophisticated hierarchy equations of motion approach~\cite{tanimura06_082001,hein12_023018} This would also give access to  effects of coherent exciton-vibrational dynamics, which have been shown to be important even in case of F\"orster transfer.~\cite{mancal13_135}

\section*{Acknowledgments}
T.P. gratefully  acknowledges financial support by  the Swedish Research Council and NanoLund.

%\bibliography{B-FD2DS}

\begin{thebibliography}{46}%
\makeatletter
\providecommand \@ifxundefined [1]{%
 \@ifx{#1\undefined}
}%
\providecommand \@ifnum [1]{%
 \ifnum #1\expandafter \@firstoftwo
 \else \expandafter \@secondoftwo
 \fi
}%
\providecommand \@ifx [1]{%
 \ifx #1\expandafter \@firstoftwo
 \else \expandafter \@secondoftwo
 \fi
}%
\providecommand \natexlab [1]{#1}%
\providecommand \enquote  [1]{``#1''}%
\providecommand \bibnamefont  [1]{#1}%
\providecommand \bibfnamefont [1]{#1}%
\providecommand \citenamefont [1]{#1}%
\providecommand \href@noop [0]{\@secondoftwo}%
\providecommand \href [0]{\begingroup \@sanitize@url \@href}%
\providecommand \@href[1]{\@@startlink{#1}\@@href}%
\providecommand \@@href[1]{\endgroup#1\@@endlink}%
\providecommand \@sanitize@url [0]{\catcode `\\12\catcode `\$12\catcode
  `\&12\catcode `\#12\catcode `\^12\catcode `\_12\catcode `\%12\relax}%
\providecommand \@@startlink[1]{}%
\providecommand \@@endlink[0]{}%
\providecommand \url  [0]{\begingroup\@sanitize@url \@url }%
\providecommand \@url [1]{\endgroup\@href {#1}{\urlprefix }}%
\providecommand \urlprefix  [0]{URL }%
\providecommand \Eprint [0]{\href }%
\providecommand \doibase [0]{http://dx.doi.org/}%
\providecommand \selectlanguage [0]{\@gobble}%
\providecommand \bibinfo  [0]{\@secondoftwo}%
\providecommand \bibfield  [0]{\@secondoftwo}%
\providecommand \translation [1]{[#1]}%
\providecommand \BibitemOpen [0]{}%
\providecommand \bibitemStop [0]{}%
\providecommand \bibitemNoStop [0]{.\EOS\space}%
\providecommand \EOS [0]{\spacefactor3000\relax}%
\providecommand \BibitemShut  [1]{\csname bibitem#1\endcsname}%
\let\auto@bib@innerbib\@empty
%</preamble>
\bibitem [{\citenamefont {May}\ and\ \citenamefont {K\"uhn}(2011)}]{may11}%
  \BibitemOpen
  \bibfield  {author} {\bibinfo {author} {\bibfnamefont {V.}~\bibnamefont
  {May}}\ and\ \bibinfo {author} {\bibfnamefont {O.}~\bibnamefont {K\"uhn}},\
  }\href@noop {} {\emph {\bibinfo {title} {Charge and {{Energy Transfer
  Dynamics}} in {{Molecular Systems}}, 3rd Revised and Enlarged Edition}}}\
  (\bibinfo  {publisher} {{Wiley-VCH}},\ \bibinfo {address} {Weinheim},\
  \bibinfo {year} {2011})\BibitemShut {NoStop}%
\bibitem [{\citenamefont {F\"orster}(1948)}]{forster48_55}%
  \BibitemOpen
  \bibfield  {author} {\bibinfo {author} {\bibfnamefont {T.}~\bibnamefont
  {F\"orster}},\ }\href@noop {} {\bibfield  {journal} {\bibinfo  {journal}
  {Ann. Physik (Leipzig)}\ }\textbf {\bibinfo {volume} {6}},\ \bibinfo {pages}
  {55} (\bibinfo {year} {1948})}\BibitemShut {NoStop}%
\bibitem [{\citenamefont {Scholes}(2003)}]{scholes03_57}%
  \BibitemOpen
  \bibfield  {author} {\bibinfo {author} {\bibfnamefont {G.~D.}\ \bibnamefont
  {Scholes}},\ }\href@noop {} {\bibfield  {journal} {\bibinfo  {journal} {Annu.
  Rev. Phys. Chem.}\ }\textbf {\bibinfo {volume} {54}},\ \bibinfo {pages} {57}
  (\bibinfo {year} {2003})}\BibitemShut {NoStop}%
\bibitem [{\citenamefont {Beljonne}\ \emph {et~al.}(2009)\citenamefont
  {Beljonne}, \citenamefont {Curutchet}, \citenamefont {Scholes},\ and\
  \citenamefont {Silbey}}]{beljonne09_6583}%
  \BibitemOpen
  \bibfield  {author} {\bibinfo {author} {\bibfnamefont {D.}~\bibnamefont
  {Beljonne}}, \bibinfo {author} {\bibfnamefont {C.}~\bibnamefont {Curutchet}},
  \bibinfo {author} {\bibfnamefont {G.~D.}\ \bibnamefont {Scholes}}, \ and\
  \bibinfo {author} {\bibfnamefont {R.~J.}\ \bibnamefont {Silbey}},\
  }\href@noop {} {\bibfield  {journal} {\bibinfo  {journal} {J. Phys. Chem. B}\
  }\textbf {\bibinfo {volume} {113}},\ \bibinfo {pages} {6583} (\bibinfo {year}
  {2009})}\BibitemShut {NoStop}%
\bibitem [{\citenamefont {{\c S}ener}\ \emph {et~al.}(2007)\citenamefont {{\c
  S}ener}, \citenamefont {Olsen}, \citenamefont {Hunter},\ and\ \citenamefont
  {Schulten}}]{sener07_15723}%
  \BibitemOpen
  \bibfield  {author} {\bibinfo {author} {\bibfnamefont {M.~K.}\ \bibnamefont
  {{\c S}ener}}, \bibinfo {author} {\bibfnamefont {J.~D.}\ \bibnamefont
  {Olsen}}, \bibinfo {author} {\bibfnamefont {C.~N.}\ \bibnamefont {Hunter}}, \
  and\ \bibinfo {author} {\bibfnamefont {K.}~\bibnamefont {Schulten}},\
  }\href@noop {} {\bibfield  {journal} {\bibinfo  {journal} {Proc. Natl. Acad.
  Sci. USA}\ }\textbf {\bibinfo {volume} {104}},\ \bibinfo {pages} {15723}
  (\bibinfo {year} {2007})}\BibitemShut {NoStop}%
\bibitem [{\citenamefont {{\c S}ener}\ \emph {et~al.}(2011)\citenamefont {{\c
  S}ener}, \citenamefont {Str\"umpfer}, \citenamefont {Hsin}, \citenamefont
  {Chandler}, \citenamefont {Scheuring}, \citenamefont {Hunter},\ and\
  \citenamefont {Schulten}}]{sener11_518}%
  \BibitemOpen
  \bibfield  {author} {\bibinfo {author} {\bibfnamefont {M.}~\bibnamefont {{\c
  S}ener}}, \bibinfo {author} {\bibfnamefont {J.}~\bibnamefont {Str\"umpfer}},
  \bibinfo {author} {\bibfnamefont {J.}~\bibnamefont {Hsin}}, \bibinfo {author}
  {\bibfnamefont {D.}~\bibnamefont {Chandler}}, \bibinfo {author}
  {\bibfnamefont {S.}~\bibnamefont {Scheuring}}, \bibinfo {author}
  {\bibfnamefont {C.~N.}\ \bibnamefont {Hunter}}, \ and\ \bibinfo {author}
  {\bibfnamefont {K.}~\bibnamefont {Schulten}},\ }\href@noop {} {\ \textbf
  {\bibinfo {volume} {12}},\ \bibinfo {pages} {518} (\bibinfo {year}
  {2011})}\BibitemShut {NoStop}%
\bibitem [{\citenamefont {Pullerits}\ \emph {et~al.}(1997)\citenamefont
  {Pullerits}, \citenamefont {Hess}, \citenamefont {Herek},\ and\ \citenamefont
  {Sundstr\"om}}]{pullerits97_10560}%
  \BibitemOpen
  \bibfield  {author} {\bibinfo {author} {\bibfnamefont {T.~o.}\ \bibnamefont
  {Pullerits}}, \bibinfo {author} {\bibfnamefont {S.}~\bibnamefont {Hess}},
  \bibinfo {author} {\bibfnamefont {J.~L.}\ \bibnamefont {Herek}}, \ and\
  \bibinfo {author} {\bibfnamefont {V.}~\bibnamefont {Sundstr\"om}},\
  }\href@noop {} {\bibfield  {journal} {\bibinfo  {journal} {J. Phys. Chem. B}\
  }\textbf {\bibinfo {volume} {101}},\ \bibinfo {pages} {10560} (\bibinfo
  {year} {1997})}\BibitemShut {NoStop}%
\bibitem [{\citenamefont {Jonas}(2003)}]{jonas03_425}%
  \BibitemOpen
  \bibfield  {author} {\bibinfo {author} {\bibfnamefont {D.~M.}\ \bibnamefont
  {Jonas}},\ }\href@noop {} {\bibfield  {journal} {\bibinfo  {journal} {Annu.
  Rev. Phys. Chem.}\ }\textbf {\bibinfo {volume} {54}},\ \bibinfo {pages} {425}
  (\bibinfo {year} {2003})}\BibitemShut {NoStop}%
\bibitem [{\citenamefont {Brixner}\ \emph {et~al.}(2005)\citenamefont
  {Brixner}, \citenamefont {Stenger}, \citenamefont {Vaswani}, \citenamefont
  {Cho}, \citenamefont {Blankenship},\ and\ \citenamefont
  {Fleming}}]{brixner05_625}%
  \BibitemOpen
  \bibfield  {author} {\bibinfo {author} {\bibfnamefont {T.}~\bibnamefont
  {Brixner}}, \bibinfo {author} {\bibfnamefont {J.}~\bibnamefont {Stenger}},
  \bibinfo {author} {\bibfnamefont {H.~M.}\ \bibnamefont {Vaswani}}, \bibinfo
  {author} {\bibfnamefont {M.}~\bibnamefont {Cho}}, \bibinfo {author}
  {\bibfnamefont {R.~E.}\ \bibnamefont {Blankenship}}, \ and\ \bibinfo {author}
  {\bibfnamefont {G.~R.}\ \bibnamefont {Fleming}},\ }\href@noop {} {\bibfield
  {journal} {\bibinfo  {journal} {Nature}\ }\textbf {\bibinfo {volume} {434}},\
  \bibinfo {pages} {625} (\bibinfo {year} {2005})}\BibitemShut {NoStop}%
\bibitem [{\citenamefont {Fuller}\ and\ \citenamefont
  {Ogilvie}(2015)}]{fuller15_667}%
  \BibitemOpen
  \bibfield  {author} {\bibinfo {author} {\bibfnamefont {F.~D.}\ \bibnamefont
  {Fuller}}\ and\ \bibinfo {author} {\bibfnamefont {J.~P.}\ \bibnamefont
  {Ogilvie}},\ }\href@noop {} {\bibfield  {journal} {\bibinfo  {journal} {Annu.
  Rev. Phys. Chem.}\ }\textbf {\bibinfo {volume} {66}},\ \bibinfo {pages} {667}
  (\bibinfo {year} {2015})}\BibitemShut {NoStop}%
\bibitem [{\citenamefont {Mueller}\ \emph {et~al.}(2018)\citenamefont
  {Mueller}, \citenamefont {Draeger}, \citenamefont {Ma}, \citenamefont
  {Hensen}, \citenamefont {Kenneweg}, \citenamefont {Pfeiffer},\ and\
  \citenamefont {Brixner}}]{mueller18_1964}%
  \BibitemOpen
  \bibfield  {author} {\bibinfo {author} {\bibfnamefont {S.}~\bibnamefont
  {Mueller}}, \bibinfo {author} {\bibfnamefont {S.}~\bibnamefont {Draeger}},
  \bibinfo {author} {\bibfnamefont {X.}~\bibnamefont {Ma}}, \bibinfo {author}
  {\bibfnamefont {M.}~\bibnamefont {Hensen}}, \bibinfo {author} {\bibfnamefont
  {T.}~\bibnamefont {Kenneweg}}, \bibinfo {author} {\bibfnamefont
  {W.}~\bibnamefont {Pfeiffer}}, \ and\ \bibinfo {author} {\bibfnamefont
  {T.}~\bibnamefont {Brixner}},\ }\href@noop {} {\bibfield  {journal} {\bibinfo
   {journal} {J. Phys. Chem. Lett.}\ }\textbf {\bibinfo {volume} {9}},\
  \bibinfo {pages} {1964} (\bibinfo {year} {2018})}\BibitemShut {NoStop}%
\bibitem [{\citenamefont {Aeschlimann}\ \emph {et~al.}(2015)\citenamefont
  {Aeschlimann}, \citenamefont {Brixner}, \citenamefont {Differt},
  \citenamefont {Heinzmann}, \citenamefont {Hensen}, \citenamefont {Kramer},
  \citenamefont {L\"ukermann}, \citenamefont {Melchior}, \citenamefont
  {Pfeiffer}, \citenamefont {Piecuch}, \citenamefont {Schneider}, \citenamefont
  {Stiebig}, \citenamefont {Str\"uber},\ and\ \citenamefont
  {Thielen}}]{aeschlimann15_663}%
  \BibitemOpen
  \bibfield  {author} {\bibinfo {author} {\bibfnamefont {M.}~\bibnamefont
  {Aeschlimann}}, \bibinfo {author} {\bibfnamefont {T.}~\bibnamefont
  {Brixner}}, \bibinfo {author} {\bibfnamefont {D.}~\bibnamefont {Differt}},
  \bibinfo {author} {\bibfnamefont {U.}~\bibnamefont {Heinzmann}}, \bibinfo
  {author} {\bibfnamefont {M.}~\bibnamefont {Hensen}}, \bibinfo {author}
  {\bibfnamefont {C.}~\bibnamefont {Kramer}}, \bibinfo {author} {\bibfnamefont
  {F.}~\bibnamefont {L\"ukermann}}, \bibinfo {author} {\bibfnamefont
  {P.}~\bibnamefont {Melchior}}, \bibinfo {author} {\bibfnamefont
  {W.}~\bibnamefont {Pfeiffer}}, \bibinfo {author} {\bibfnamefont
  {M.}~\bibnamefont {Piecuch}}, \bibinfo {author} {\bibfnamefont
  {C.}~\bibnamefont {Schneider}}, \bibinfo {author} {\bibfnamefont
  {H.}~\bibnamefont {Stiebig}}, \bibinfo {author} {\bibfnamefont
  {C.}~\bibnamefont {Str\"uber}}, \ and\ \bibinfo {author} {\bibfnamefont
  {P.}~\bibnamefont {Thielen}},\ }\href@noop {} {\bibfield  {journal} {\bibinfo
   {journal} {Nature Photonics}\ }\textbf {\bibinfo {volume} {9}},\ \bibinfo
  {pages} {663} (\bibinfo {year} {2015})}\BibitemShut {NoStop}%
\bibitem [{\citenamefont {Nardin}\ \emph {et~al.}(2013)\citenamefont {Nardin},
  \citenamefont {Autry}, \citenamefont {Silverman},\ and\ \citenamefont
  {Cundiff}}]{nardin13_28617}%
  \BibitemOpen
  \bibfield  {author} {\bibinfo {author} {\bibfnamefont {G.}~\bibnamefont
  {Nardin}}, \bibinfo {author} {\bibfnamefont {T.~M.}\ \bibnamefont {Autry}},
  \bibinfo {author} {\bibfnamefont {K.~L.}\ \bibnamefont {Silverman}}, \ and\
  \bibinfo {author} {\bibfnamefont {S.~T.}\ \bibnamefont {Cundiff}},\
  }\href@noop {} {\bibfield  {journal} {\bibinfo  {journal} {Opt. Express}\
  }\textbf {\bibinfo {volume} {21}},\ \bibinfo {pages} {28617} (\bibinfo {year}
  {2013})}\BibitemShut {NoStop}%
\bibitem [{\citenamefont {Karki}\ \emph {et~al.}(2014)\citenamefont {Karki},
  \citenamefont {Widom}, \citenamefont {Seibt}, \citenamefont {Moody},
  \citenamefont {Lonergan}, \citenamefont {Pullerits},\ and\ \citenamefont
  {Marcus}}]{karki14_5869}%
  \BibitemOpen
  \bibfield  {author} {\bibinfo {author} {\bibfnamefont {K.~J.}\ \bibnamefont
  {Karki}}, \bibinfo {author} {\bibfnamefont {J.~R.}\ \bibnamefont {Widom}},
  \bibinfo {author} {\bibfnamefont {J.}~\bibnamefont {Seibt}}, \bibinfo
  {author} {\bibfnamefont {I.}~\bibnamefont {Moody}}, \bibinfo {author}
  {\bibfnamefont {M.~C.}\ \bibnamefont {Lonergan}}, \bibinfo {author}
  {\bibfnamefont {T.~o.}\ \bibnamefont {Pullerits}}, \ and\ \bibinfo {author}
  {\bibfnamefont {A.~H.}\ \bibnamefont {Marcus}},\ }\href@noop {} {\bibfield
  {journal} {\bibinfo  {journal} {Nature Commun.}\ }\textbf {\bibinfo {volume}
  {5}},\ \bibinfo {pages} {5869} (\bibinfo {year} {2014})}\BibitemShut
  {NoStop}%
\bibitem [{\citenamefont {Lott}\ \emph {et~al.}(2011)\citenamefont {Lott},
  \citenamefont {{Perdomo-Ortiz}}, \citenamefont {Utterback}, \citenamefont
  {Widom}, \citenamefont {{Aspuru-Guzik}},\ and\ \citenamefont
  {Marcus}}]{lott11_16521}%
  \BibitemOpen
  \bibfield  {author} {\bibinfo {author} {\bibfnamefont {G.~A.}\ \bibnamefont
  {Lott}}, \bibinfo {author} {\bibfnamefont {A.}~\bibnamefont
  {{Perdomo-Ortiz}}}, \bibinfo {author} {\bibfnamefont {J.~K.}\ \bibnamefont
  {Utterback}}, \bibinfo {author} {\bibfnamefont {J.~R.}\ \bibnamefont
  {Widom}}, \bibinfo {author} {\bibfnamefont {A.}~\bibnamefont
  {{Aspuru-Guzik}}}, \ and\ \bibinfo {author} {\bibfnamefont {A.~H.}\
  \bibnamefont {Marcus}},\ }\href@noop {} {\bibfield  {journal} {\bibinfo
  {journal} {Proc. Natl. Acad. Sci. USA}\ }\textbf {\bibinfo {volume} {108}},\
  \bibinfo {pages} {16521} (\bibinfo {year} {2011})}\BibitemShut {NoStop}%
\bibitem [{\citenamefont {Tiwari}\ \emph {et~al.}(2018)\citenamefont {Tiwari},
  \citenamefont {Matutes}, \citenamefont {Yu}, \citenamefont {Ptaszek},
  \citenamefont {Bocian}, \citenamefont {Holten}, \citenamefont {Kirmaier},
  \citenamefont {Konar},\ and\ \citenamefont {Ogilvie}}]{tiwari18_}%
  \BibitemOpen
  \bibfield  {author} {\bibinfo {author} {\bibfnamefont {V.}~\bibnamefont
  {Tiwari}}, \bibinfo {author} {\bibfnamefont {Y.~A.}\ \bibnamefont {Matutes}},
  \bibinfo {author} {\bibfnamefont {Z.}~\bibnamefont {Yu}}, \bibinfo {author}
  {\bibfnamefont {M.}~\bibnamefont {Ptaszek}}, \bibinfo {author} {\bibfnamefont
  {D.~F.}\ \bibnamefont {Bocian}}, \bibinfo {author} {\bibfnamefont
  {D.}~\bibnamefont {Holten}}, \bibinfo {author} {\bibfnamefont
  {C.}~\bibnamefont {Kirmaier}}, \bibinfo {author} {\bibfnamefont
  {A.}~\bibnamefont {Konar}}, \ and\ \bibinfo {author} {\bibfnamefont {J.~P.}\
  \bibnamefont {Ogilvie}},\ }\href@noop {} {\bibfield  {journal} {\bibinfo
  {journal} {arXiv:1806.00896 [physics.chem-ph]}\ } (\bibinfo {year}
  {2018})}\BibitemShut {NoStop}%
\bibitem [{\citenamefont {Karki}\ \emph {et~al.}(2018)\citenamefont {Karki},
  \citenamefont {Chen}, \citenamefont {Sakurai}, \citenamefont {Shi},
  \citenamefont {Gardiner}, \citenamefont {K\"uhn}, \citenamefont {Cogdell},\
  and\ \citenamefont {Pullerits}}]{karki18_}%
  \BibitemOpen
  \bibfield  {author} {\bibinfo {author} {\bibfnamefont {K.~J.}\ \bibnamefont
  {Karki}}, \bibinfo {author} {\bibfnamefont {J.}~\bibnamefont {Chen}},
  \bibinfo {author} {\bibfnamefont {A.}~\bibnamefont {Sakurai}}, \bibinfo
  {author} {\bibfnamefont {Q.}~\bibnamefont {Shi}}, \bibinfo {author}
  {\bibfnamefont {A.~T.}\ \bibnamefont {Gardiner}}, \bibinfo {author}
  {\bibfnamefont {O.}~\bibnamefont {K\"uhn}}, \bibinfo {author} {\bibfnamefont
  {R.~J.}\ \bibnamefont {Cogdell}}, \ and\ \bibinfo {author} {\bibfnamefont
  {T.}~\bibnamefont {Pullerits}},\ }\href@noop {} {\bibfield  {journal}
  {\bibinfo  {journal} {arXiv:1804.04840 [physics.chem-ph]}\ } (\bibinfo {year}
  {2018})}\BibitemShut {NoStop}%
\bibitem [{\citenamefont {Abramavicius}\ \emph {et~al.}(2009)\citenamefont
  {Abramavicius}, \citenamefont {Palmieri}, \citenamefont {Voronine},
  \citenamefont {Sanda},\ and\ \citenamefont {Mukamel}}]{abramavicius09_2350}%
  \BibitemOpen
  \bibfield  {author} {\bibinfo {author} {\bibfnamefont {D.}~\bibnamefont
  {Abramavicius}}, \bibinfo {author} {\bibfnamefont {B.}~\bibnamefont
  {Palmieri}}, \bibinfo {author} {\bibfnamefont {D.~V.}\ \bibnamefont
  {Voronine}}, \bibinfo {author} {\bibfnamefont {F.}~\bibnamefont {Sanda}}, \
  and\ \bibinfo {author} {\bibfnamefont {S.}~\bibnamefont {Mukamel}},\
  }\href@noop {} {\bibfield  {journal} {\bibinfo  {journal} {Chem. Rev.}\
  }\textbf {\bibinfo {volume} {109}},\ \bibinfo {pages} {2350} (\bibinfo {year}
  {2009})}\BibitemShut {NoStop}%
\bibitem [{\citenamefont {Mukamel}(1995)}]{mukamel95_}%
  \BibitemOpen
  \bibfield  {author} {\bibinfo {author} {\bibfnamefont {S.}~\bibnamefont
  {Mukamel}},\ }\href@noop {} {\emph {\bibinfo {title} {Principles of Nonlinear
  Optical Spectroscopy}}}\ (\bibinfo  {publisher} {{Oxford University Press}},\
  \bibinfo {address} {Oxford},\ \bibinfo {year} {1995})\BibitemShut {NoStop}%
\bibitem [{\citenamefont {Thyrhaug}\ \emph {et~al.}(2016)\citenamefont
  {Thyrhaug}, \citenamefont {Zidek}, \citenamefont {Dostal}, \citenamefont
  {Bina},\ and\ \citenamefont {Zigmantas}}]{thyrhaug16_1653}%
  \BibitemOpen
  \bibfield  {author} {\bibinfo {author} {\bibfnamefont {E.}~\bibnamefont
  {Thyrhaug}}, \bibinfo {author} {\bibfnamefont {K.}~\bibnamefont {Zidek}},
  \bibinfo {author} {\bibfnamefont {J.}~\bibnamefont {Dostal}}, \bibinfo
  {author} {\bibfnamefont {D.}~\bibnamefont {Bina}}, \ and\ \bibinfo {author}
  {\bibfnamefont {D.}~\bibnamefont {Zigmantas}},\ }\href@noop {} {\bibfield
  {journal} {\bibinfo  {journal} {J. Phys. Chem. Lett.}\ }\textbf {\bibinfo
  {volume} {7}},\ \bibinfo {pages} {1653} (\bibinfo {year} {2016})}\BibitemShut
  {NoStop}%
\bibitem [{\citenamefont {Harel}\ and\ \citenamefont
  {Engel}(2012)}]{harel12_706}%
  \BibitemOpen
  \bibfield  {author} {\bibinfo {author} {\bibfnamefont {E.}~\bibnamefont
  {Harel}}\ and\ \bibinfo {author} {\bibfnamefont {G.~S.}\ \bibnamefont
  {Engel}},\ }\href@noop {} {\bibfield  {journal} {\bibinfo  {journal} {Proc.
  Natl. Acad. Sci. USA}\ }\textbf {\bibinfo {volume} {109}},\ \bibinfo {pages}
  {706} (\bibinfo {year} {2012})}\BibitemShut {NoStop}%
\bibitem [{\citenamefont {Schr\"oter}\ \emph {et~al.}(2018)\citenamefont
  {Schr\"oter}, \citenamefont {Alcocer}, \citenamefont {Cogdell}, \citenamefont
  {K\"uhn},\ and\ \citenamefont {Zigmantas}}]{schroter18_1340}%
  \BibitemOpen
  \bibfield  {author} {\bibinfo {author} {\bibfnamefont {M.}~\bibnamefont
  {Schr\"oter}}, \bibinfo {author} {\bibfnamefont {M.~J.~P.}\ \bibnamefont
  {Alcocer}}, \bibinfo {author} {\bibfnamefont {R.~J.}\ \bibnamefont
  {Cogdell}}, \bibinfo {author} {\bibfnamefont {O.}~\bibnamefont {K\"uhn}}, \
  and\ \bibinfo {author} {\bibfnamefont {D.}~\bibnamefont {Zigmantas}},\
  }\href@noop {} {\bibfield  {journal} {\bibinfo  {journal} {J. Phys. Chem.
  Lett.}\ }\textbf {\bibinfo {volume} {9}},\ \bibinfo {pages} {1340} (\bibinfo
  {year} {2018})}\BibitemShut {NoStop}%
\bibitem [{\citenamefont {Balevicius}, \citenamefont {Valkunas},\ and\
  \citenamefont {Abramavicius}(2015)}]{balevicius15_74101}%
  \BibitemOpen
  \bibfield  {author} {\bibinfo {author} {\bibfnamefont {V.}~\bibnamefont
  {Balevicius}}, \bibinfo {author} {\bibfnamefont {L.}~\bibnamefont
  {Valkunas}}, \ and\ \bibinfo {author} {\bibfnamefont {D.}~\bibnamefont
  {Abramavicius}},\ }\href@noop {} {\bibfield  {journal} {\bibinfo  {journal}
  {J. Chem. Phys.}\ }\textbf {\bibinfo {volume} {143}},\ \bibinfo {pages}
  {074101} (\bibinfo {year} {2015})}\BibitemShut {NoStop}%
\bibitem [{\citenamefont {K\"uhn}, \citenamefont {Chernyak},\ and\
  \citenamefont {Mukamel}(1996)}]{kuhn96_8586}%
  \BibitemOpen
  \bibfield  {author} {\bibinfo {author} {\bibfnamefont {O.}~\bibnamefont
  {K\"uhn}}, \bibinfo {author} {\bibfnamefont {V.}~\bibnamefont {Chernyak}}, \
  and\ \bibinfo {author} {\bibfnamefont {S.}~\bibnamefont {Mukamel}},\
  }\href@noop {} {\bibfield  {journal} {\bibinfo  {journal} {J. Chem. Phys.}\
  }\textbf {\bibinfo {volume} {105}},\ \bibinfo {pages} {8586} (\bibinfo {year}
  {1996})}\BibitemShut {NoStop}%
\bibitem [{\citenamefont {K\"uhn}\ and\ \citenamefont
  {Mukamel}(1997)}]{kuhn97_809}%
  \BibitemOpen
  \bibfield  {author} {\bibinfo {author} {\bibfnamefont {O.}~\bibnamefont
  {K\"uhn}}\ and\ \bibinfo {author} {\bibfnamefont {S.}~\bibnamefont
  {Mukamel}},\ }\href@noop {} {\bibfield  {journal} {\bibinfo  {journal} {J.
  Phys. Chem. B}\ }\textbf {\bibinfo {volume} {101}},\ \bibinfo {pages} {809}
  (\bibinfo {year} {1997})}\BibitemShut {NoStop}%
\bibitem [{\citenamefont {Renger}\ and\ \citenamefont
  {May}(1997)}]{renger97_3406}%
  \BibitemOpen
  \bibfield  {author} {\bibinfo {author} {\bibfnamefont {T.}~\bibnamefont
  {Renger}}\ and\ \bibinfo {author} {\bibfnamefont {V.}~\bibnamefont {May}},\
  }\href@noop {} {\bibfield  {journal} {\bibinfo  {journal} {Phys. Rev. Lett.}\
  }\textbf {\bibinfo {volume} {78}},\ \bibinfo {pages} {3406} (\bibinfo {year}
  {1997})}\BibitemShut {NoStop}%
\bibitem [{\citenamefont {Br\"uggemann}\ and\ \citenamefont
  {May}(2003)}]{bruggemann03_746}%
  \BibitemOpen
  \bibfield  {author} {\bibinfo {author} {\bibfnamefont {B.}~\bibnamefont
  {Br\"uggemann}}\ and\ \bibinfo {author} {\bibfnamefont {V.}~\bibnamefont
  {May}},\ }\href@noop {} {\bibfield  {journal} {\bibinfo  {journal} {J. Chem.
  Phys.}\ }\textbf {\bibinfo {volume} {118}},\ \bibinfo {pages} {746} (\bibinfo
  {year} {2003})}\BibitemShut {NoStop}%
\bibitem [{\citenamefont {Yan}\ and\ \citenamefont
  {K\"uhn}(2012)}]{yan12_105004}%
  \BibitemOpen
  \bibfield  {author} {\bibinfo {author} {\bibfnamefont {Y.}~\bibnamefont
  {Yan}}\ and\ \bibinfo {author} {\bibfnamefont {O.}~\bibnamefont {K\"uhn}},\
  }\href@noop {} {\bibfield  {journal} {\bibinfo  {journal} {New J. Phys.}\
  }\textbf {\bibinfo {volume} {14}},\ \bibinfo {pages} {105004} (\bibinfo
  {year} {2012})}\BibitemShut {NoStop}%
\bibitem [{\citenamefont {May}(2009)}]{may09_10086}%
  \BibitemOpen
  \bibfield  {author} {\bibinfo {author} {\bibfnamefont {V.}~\bibnamefont
  {May}},\ }\href@noop {} {\bibfield  {journal} {\bibinfo  {journal} {Dalton
  Trans.}\ ,\ \bibinfo {pages} {10086}} (\bibinfo {year} {2009})}\BibitemShut
  {NoStop}%
\bibitem [{\citenamefont {Damtie}\ \emph {et~al.}(2017)\citenamefont {Damtie},
  \citenamefont {Wacker}, \citenamefont {Pullerits},\ and\ \citenamefont
  {Karki}}]{damtie17_053830}%
  \BibitemOpen
  \bibfield  {author} {\bibinfo {author} {\bibfnamefont {F.~A.}\ \bibnamefont
  {Damtie}}, \bibinfo {author} {\bibfnamefont {A.}~\bibnamefont {Wacker}},
  \bibinfo {author} {\bibfnamefont {T.~o.}\ \bibnamefont {Pullerits}}, \ and\
  \bibinfo {author} {\bibfnamefont {K.~J.}\ \bibnamefont {Karki}},\ }\href@noop
  {} {\bibfield  {journal} {\bibinfo  {journal} {Phys. Rev. A}\ }\textbf
  {\bibinfo {volume} {96}},\ \bibinfo {pages} {053830} (\bibinfo {year}
  {2017})}\BibitemShut {NoStop}%
\bibitem [{\citenamefont {Pullerits}\ and\ \citenamefont
  {Sundstr\"om}(1996)}]{pullerits96_381}%
  \BibitemOpen
  \bibfield  {author} {\bibinfo {author} {\bibfnamefont {T.}~\bibnamefont
  {Pullerits}}\ and\ \bibinfo {author} {\bibfnamefont {V.}~\bibnamefont
  {Sundstr\"om}},\ }\href@noop {} {\bibfield  {journal} {\bibinfo  {journal}
  {Acc. Chem. Res.}\ }\textbf {\bibinfo {volume} {29}},\ \bibinfo {pages} {381}
  (\bibinfo {year} {1996})}\BibitemShut {NoStop}%
\bibitem [{\citenamefont {K\"uhn}\ and\ \citenamefont
  {Sundstr\"om}(1997{\natexlab{a}})}]{kuhn97_4154}%
  \BibitemOpen
  \bibfield  {author} {\bibinfo {author} {\bibfnamefont {O.}~\bibnamefont
  {K\"uhn}}\ and\ \bibinfo {author} {\bibfnamefont {V.}~\bibnamefont
  {Sundstr\"om}},\ }\href@noop {} {\bibfield  {journal} {\bibinfo  {journal}
  {J. Chem. Phys.}\ }\textbf {\bibinfo {volume} {107}},\ \bibinfo {pages}
  {4154} (\bibinfo {year} {1997}{\natexlab{a}})}\BibitemShut {NoStop}%
\bibitem [{\citenamefont {Ambrosek}\ \emph {et~al.}(2011)\citenamefont
  {Ambrosek}, \citenamefont {Marciniak}, \citenamefont {Lochbrunner},
  \citenamefont {Tatchen}, \citenamefont {Li}, \citenamefont {W\"urthner},\
  and\ \citenamefont {K\"uhn}}]{ambrosek11_17649}%
  \BibitemOpen
  \bibfield  {author} {\bibinfo {author} {\bibfnamefont {D.}~\bibnamefont
  {Ambrosek}}, \bibinfo {author} {\bibfnamefont {H.}~\bibnamefont {Marciniak}},
  \bibinfo {author} {\bibfnamefont {S.}~\bibnamefont {Lochbrunner}}, \bibinfo
  {author} {\bibfnamefont {J.}~\bibnamefont {Tatchen}}, \bibinfo {author}
  {\bibfnamefont {X.-Q.}\ \bibnamefont {Li}}, \bibinfo {author} {\bibfnamefont
  {F.}~\bibnamefont {W\"urthner}}, \ and\ \bibinfo {author} {\bibfnamefont
  {O.}~\bibnamefont {K\"uhn}},\ }\href@noop {} {\bibfield  {journal} {\bibinfo
  {journal} {Phys. Chem. Chem. Phys.}\ }\textbf {\bibinfo {volume} {13}},\
  \bibinfo {pages} {17649} (\bibinfo {year} {2011})}\BibitemShut {NoStop}%
\bibitem [{\citenamefont {Kosumi}\ \emph {et~al.}(2011)\citenamefont {Kosumi},
  \citenamefont {Maruta}, \citenamefont {Fujii}, \citenamefont {Kanemoto},
  \citenamefont {Sugisaki},\ and\ \citenamefont {Hashimoto}}]{kosumi11_92}%
  \BibitemOpen
  \bibfield  {author} {\bibinfo {author} {\bibfnamefont {D.}~\bibnamefont
  {Kosumi}}, \bibinfo {author} {\bibfnamefont {S.}~\bibnamefont {Maruta}},
  \bibinfo {author} {\bibfnamefont {R.}~\bibnamefont {Fujii}}, \bibinfo
  {author} {\bibfnamefont {K.}~\bibnamefont {Kanemoto}}, \bibinfo {author}
  {\bibfnamefont {M.}~\bibnamefont {Sugisaki}}, \ and\ \bibinfo {author}
  {\bibfnamefont {H.}~\bibnamefont {Hashimoto}},\ }\href@noop {} {\bibfield
  {journal} {\bibinfo  {journal} {phys. stat. sol. C}\ }\textbf {\bibinfo
  {volume} {8}},\ \bibinfo {pages} {92} (\bibinfo {year} {2011})}\BibitemShut
  {NoStop}%
\bibitem [{\citenamefont {Br\"uggemann}, \citenamefont {Christensson},\ and\
  \citenamefont {Pullerits}(2009)}]{bruggemann09_140}%
  \BibitemOpen
  \bibfield  {author} {\bibinfo {author} {\bibfnamefont {B.}~\bibnamefont
  {Br\"uggemann}}, \bibinfo {author} {\bibfnamefont {N.}~\bibnamefont
  {Christensson}}, \ and\ \bibinfo {author} {\bibfnamefont {T.}~\bibnamefont
  {Pullerits}},\ }\href@noop {} {\bibfield  {journal} {\bibinfo  {journal}
  {Chem. Phys.}\ }\textbf {\bibinfo {volume} {357}},\ \bibinfo {pages} {140}
  (\bibinfo {year} {2009})}\BibitemShut {NoStop}%
\bibitem [{\citenamefont {Hader}\ \emph {et~al.}(2017)\citenamefont {Hader},
  \citenamefont {Consani}, \citenamefont {Brixner},\ and\ \citenamefont
  {Engel}}]{hader17_31989}%
  \BibitemOpen
  \bibfield  {author} {\bibinfo {author} {\bibfnamefont {K.}~\bibnamefont
  {Hader}}, \bibinfo {author} {\bibfnamefont {C.}~\bibnamefont {Consani}},
  \bibinfo {author} {\bibfnamefont {T.}~\bibnamefont {Brixner}}, \ and\
  \bibinfo {author} {\bibfnamefont {V.}~\bibnamefont {Engel}},\ }\href@noop {}
  {\bibfield  {journal} {\bibinfo  {journal} {Phys. Chem. Chem. Phys.}\
  }\textbf {\bibinfo {volume} {19}},\ \bibinfo {pages} {31989} (\bibinfo {year}
  {2017})}\BibitemShut {NoStop}%
\bibitem [{\citenamefont {Dost\'al}\ \emph {et~al.}(2018)\citenamefont
  {Dost\'al}, \citenamefont {Fennel}, \citenamefont {Koch}, \citenamefont
  {Herbst}, \citenamefont {W\"urthner},\ and\ \citenamefont
  {Brixner}}]{dostal18_2466}%
  \BibitemOpen
  \bibfield  {author} {\bibinfo {author} {\bibfnamefont {J.}~\bibnamefont
  {Dost\'al}}, \bibinfo {author} {\bibfnamefont {F.}~\bibnamefont {Fennel}},
  \bibinfo {author} {\bibfnamefont {F.}~\bibnamefont {Koch}}, \bibinfo {author}
  {\bibfnamefont {S.}~\bibnamefont {Herbst}}, \bibinfo {author} {\bibfnamefont
  {F.}~\bibnamefont {W\"urthner}}, \ and\ \bibinfo {author} {\bibfnamefont
  {T.}~\bibnamefont {Brixner}},\ }\href@noop {} {\bibfield  {journal} {\bibinfo
   {journal} {Nature Commun.}\ }\textbf {\bibinfo {volume} {9}},\ \bibinfo
  {pages} {2466} (\bibinfo {year} {2018})}\BibitemShut {NoStop}%
\bibitem [{\citenamefont {K\"uhn}\ and\ \citenamefont
  {Sundstr\"om}(1997{\natexlab{b}})}]{kuhn97_3432}%
  \BibitemOpen
  \bibfield  {author} {\bibinfo {author} {\bibfnamefont {O.}~\bibnamefont
  {K\"uhn}}\ and\ \bibinfo {author} {\bibfnamefont {V.}~\bibnamefont
  {Sundstr\"om}},\ }\href@noop {} {\bibfield  {journal} {\bibinfo  {journal}
  {J. Phys. Chem. B}\ }\textbf {\bibinfo {volume} {101}},\ \bibinfo {pages}
  {3432} (\bibinfo {year} {1997}{\natexlab{b}})}\BibitemShut {NoStop}%
\bibitem [{\citenamefont {Woller}, \citenamefont {Hannestad},\ and\
  \citenamefont {Albinsson}(2013)}]{woller13_2759}%
  \BibitemOpen
  \bibfield  {author} {\bibinfo {author} {\bibfnamefont {J.~G.}\ \bibnamefont
  {Woller}}, \bibinfo {author} {\bibfnamefont {J.~K.}\ \bibnamefont
  {Hannestad}}, \ and\ \bibinfo {author} {\bibfnamefont {B.}~\bibnamefont
  {Albinsson}},\ }\href@noop {} {\bibfield  {journal} {\bibinfo  {journal} {J.
  Am. Chem. Soc.}\ }\textbf {\bibinfo {volume} {135}},\ \bibinfo {pages} {2759}
  (\bibinfo {year} {2013})}\BibitemShut {NoStop}%
\bibitem [{\citenamefont {Kramer}\ \emph {et~al.}(2018)\citenamefont {Kramer},
  \citenamefont {Noack}, \citenamefont {Reimers}, \citenamefont {Reinefeld},
  \citenamefont {Rodríguez},\ and\ \citenamefont {Yin}}]{kramer18_}%
  \BibitemOpen
  \bibfield  {author} {\bibinfo {author} {\bibfnamefont {T.}~\bibnamefont
  {Kramer}}, \bibinfo {author} {\bibfnamefont {M.}~\bibnamefont {Noack}},
  \bibinfo {author} {\bibfnamefont {J.~R.}\ \bibnamefont {Reimers}}, \bibinfo
  {author} {\bibfnamefont {A.}~\bibnamefont {Reinefeld}}, \bibinfo {author}
  {\bibfnamefont {M.}~\bibnamefont {Rodríguez}}, \ and\ \bibinfo {author}
  {\bibfnamefont {S.}~\bibnamefont {Yin}},\ }\href@noop {} {\bibfield
  {journal} {\bibinfo  {journal} {Chem. Phys.}\ } (\bibinfo {year}
  {2018})}\BibitemShut {NoStop}%
\bibitem [{\citenamefont {{Caycedo-Soler}}\ \emph {et~al.}(2017)\citenamefont
  {{Caycedo-Soler}}, \citenamefont {Schroeder}, \citenamefont {Autenrieth},
  \citenamefont {Pick}, \citenamefont {Ghosh}, \citenamefont {Huelga},\ and\
  \citenamefont {Plenio}}]{caycedo-soler17_6015}%
  \BibitemOpen
  \bibfield  {author} {\bibinfo {author} {\bibfnamefont {F.}~\bibnamefont
  {{Caycedo-Soler}}}, \bibinfo {author} {\bibfnamefont {C.~A.}\ \bibnamefont
  {Schroeder}}, \bibinfo {author} {\bibfnamefont {C.}~\bibnamefont
  {Autenrieth}}, \bibinfo {author} {\bibfnamefont {A.}~\bibnamefont {Pick}},
  \bibinfo {author} {\bibfnamefont {R.}~\bibnamefont {Ghosh}}, \bibinfo
  {author} {\bibfnamefont {S.~F.}\ \bibnamefont {Huelga}}, \ and\ \bibinfo
  {author} {\bibfnamefont {M.~B.}\ \bibnamefont {Plenio}},\ }\href@noop {}
  {\bibfield  {journal} {\bibinfo  {journal} {J. Phys. Chem. Lett.}\ }\textbf
  {\bibinfo {volume} {8}},\ \bibinfo {pages} {6015} (\bibinfo {year}
  {2017})}\BibitemShut {NoStop}%
\bibitem [{\citenamefont {Br\"uggemann}\ \emph {et~al.}(2001)\citenamefont
  {Br\"uggemann}, \citenamefont {Herek}, \citenamefont {Sundstr\"om},
  \citenamefont {Pullerits},\ and\ \citenamefont {May}}]{bruggemann01_11391}%
  \BibitemOpen
  \bibfield  {author} {\bibinfo {author} {\bibfnamefont {B.}~\bibnamefont
  {Br\"uggemann}}, \bibinfo {author} {\bibfnamefont {J.}~\bibnamefont {Herek}},
  \bibinfo {author} {\bibfnamefont {V.}~\bibnamefont {Sundstr\"om}}, \bibinfo
  {author} {\bibfnamefont {T.}~\bibnamefont {Pullerits}}, \ and\ \bibinfo
  {author} {\bibfnamefont {V.}~\bibnamefont {May}},\ }\href@noop {} {\bibfield
  {journal} {\bibinfo  {journal} {J. Phys. Chem. B}\ }\textbf {\bibinfo
  {volume} {105}},\ \bibinfo {pages} {11391} (\bibinfo {year}
  {2001})}\BibitemShut {NoStop}%
\bibitem [{\citenamefont {K\"uhn}\ and\ \citenamefont
  {Tanimura}(2003)}]{kuhn03_2155}%
  \BibitemOpen
  \bibfield  {author} {\bibinfo {author} {\bibfnamefont {O.}~\bibnamefont
  {K\"uhn}}\ and\ \bibinfo {author} {\bibfnamefont {Y.}~\bibnamefont
  {Tanimura}},\ }\href@noop {} {\bibfield  {journal} {\bibinfo  {journal} {J.
  Chem. Phys.}\ }\textbf {\bibinfo {volume} {119}},\ \bibinfo {pages} {2155}
  (\bibinfo {year} {2003})}\BibitemShut {NoStop}%
\bibitem [{\citenamefont {Tanimura}(2006)}]{tanimura06_082001}%
  \BibitemOpen
  \bibfield  {author} {\bibinfo {author} {\bibfnamefont {Y.}~\bibnamefont
  {Tanimura}},\ }\href@noop {} {\bibfield  {journal} {\bibinfo  {journal} {J.
  Phys. Soc. Japan}\ }\textbf {\bibinfo {volume} {75}},\ \bibinfo {pages}
  {082001} (\bibinfo {year} {2006})}\BibitemShut {NoStop}%
\bibitem [{\citenamefont {Hein}\ \emph {et~al.}(2012)\citenamefont {Hein},
  \citenamefont {Kreisbeck}, \citenamefont {Kramer},\ and\ \citenamefont
  {Rodriguez}}]{hein12_023018}%
  \BibitemOpen
  \bibfield  {author} {\bibinfo {author} {\bibfnamefont {B.}~\bibnamefont
  {Hein}}, \bibinfo {author} {\bibfnamefont {C.}~\bibnamefont {Kreisbeck}},
  \bibinfo {author} {\bibfnamefont {T.}~\bibnamefont {Kramer}}, \ and\ \bibinfo
  {author} {\bibfnamefont {M.}~\bibnamefont {Rodriguez}},\ }\href@noop {}
  {\bibfield  {journal} {\bibinfo  {journal} {New J. Phys.}\ }\textbf {\bibinfo
  {volume} {14}},\ \bibinfo {pages} {023018} (\bibinfo {year}
  {2012})}\BibitemShut {NoStop}%
\bibitem [{\citenamefont {Man{\v c}al}\ \emph {et~al.}(2013)\citenamefont
  {Man{\v c}al}, \citenamefont {Dost\'al}, \citenamefont {P{\v s}en{\v
  c}\'ik},\ and\ \citenamefont {Zigmantas}}]{mancal13_135}%
  \BibitemOpen
  \bibfield  {author} {\bibinfo {author} {\bibfnamefont {T.}~\bibnamefont
  {Man{\v c}al}}, \bibinfo {author} {\bibfnamefont {J.}~\bibnamefont
  {Dost\'al}}, \bibinfo {author} {\bibfnamefont {J.}~\bibnamefont {P{\v s}en{\v
  c}\'ik}}, \ and\ \bibinfo {author} {\bibfnamefont {D.}~\bibnamefont
  {Zigmantas}},\ }\href@noop {} {\bibfield  {journal} {\bibinfo  {journal}
  {Can. J. Chem.}\ }\textbf {\bibinfo {volume} {92}},\ \bibinfo {pages} {135}
  (\bibinfo {year} {2013})}\BibitemShut {NoStop}%
\end{thebibliography}

\end{document}